# Nonlinear waves in stratified Taylor–Couette flow. Part 1. Layer formation.

Colin Leclercq[1]†, Jamie L. Partridge[2], Pierre Augier[2], Colm-Cille P. Caulfield[2], Stuart B. Dalziel[2] and Paul F. Linden[2]

[1]School of Mathematics, University of Bristol, University Walk, Bristol BS8 1TW, UK

[2]Department of Applied Mathematics and Theoretical Physics, Centre for Mathematical Sciences, Wilberforce Road, Cambridge CB3 0WA, UK

This paper is the first part of a two-fold study of mixing, i.e. the formation of layers and upwelling of buoyancy, in axially stratified Taylor–Couette flow, with fixed outer cylinder. Using linear analysis and direct numerical simulation, we show the critical role played by non-axisymmetric instability modes, despite the fact that the flow is centrifugally unstable in the sense of Rayleigh's criterion. Interactions between helical modes of opposite handedness leads to the formation of nonlinear coherent structures: (mixed)-ribbons and (mixed)-cross-spirals. These give birth to complex density interface patterns, seemingly appearing and disappearing periodically as the coherent structure slowly rotates around the annulus. These coherent structures seem to be responsible for the formation of layers reported in a recent experiment by Oglethorpe *et al.* (2013). We distinguish 'dynamic layering', instantaneous, localized and caused by the vortical motions, from 'static layering' corresponding to the formation of a 'staircase profile' in the adiabatically sorted background density. The latter only occurs at large enough Schmidt number, revealing the significant impact of the Schmidt number in the layering process.

**Key words:**

## 1. Introduction

Stratified Taylor–Couette flow (STC) is an archetype of wall-bounded stratified shear flow, with stratification perpendicular to shear, and allows to probe fundamental processes potentially relevant to geophysical fluid dynamics. A common feature of these flows is the spontaneous formation of layered structures, sometimes called 'pancake' in the context of stratified turbulence (Riley & Lelong 2000). In this paper, we investigate the physical mechanisms responsible for the formation of density patterns in axially stratified Taylor–Couette (STC) flow, revisiting Oglethorpe *et al.* (2013)'s recent experiments with linear stability analysis, direct numerical simulations, and a few additionnal experiments. In a second paper (Leclercq *et al.* ????b), the origin of the buoyancy flux is analysed in more detail.

The first experiment focusing on the formation of layers in STC with a fixed outer cylinder were performed by Boubnov *et al.* (1995) and later revisited by Caton *et al.* (1999, 2000). The modelling effort in both studies relies on the use of linear stability analysis applied to axisymmetric perturbations. The authors also recovered the scales obtained from linear analysis using Bernoulli's theorem, implicitely implying stationarity and axisymmetry. The authors admitted that the onset of instability was actually non-axisymmetric in the experiment, but attributed this behaviour to a secondary instability of an already bifurcated axisymmetric oscillatory flow. Hua *et al.* (1997) reproduced some of the experiments of Boubnov *et al.* (1995) numerically and also found that the axial scale selection was primarily due to the axisymmetric part of the flow. Caton *et al.* (1999, 2000) later proposed that the secondary bifurcation to non-axisymmetry was a global one, involving a saddle-node bifurcation of the non-axisymmetric branch very close to the (primary) axisymmetric instability threshold. All these results were obtained in a small-gap apparatus with a radius ratio of $\eta := r_i/r_o = 0.769$ ($r_i$ and $r_o$ denoting the inner and outer cylinders) and at low Reynolds numbers $Re := r_i \Omega \Delta r / \nu < 1800$, where $\Delta r := r_o - r_i$ is the gap and $\nu$ is the kinematic viscosity. More recently, Oglethorpe *et al.* (2013) extended the analysis of STC to the large-gap $\eta \in \{0.195, 0.389, 0.584\}$, large-$Re = O(10^4)$ regime, aiming to reach an asymptotic state of strongly stratified turbulence. They also observed the formation of layers separated by sharp density interfaces, which they attributed to the 'Philipps mechanism' rather than to the linear mechanism of Boubnov *et al.* (1995). They were however able to use the scaling law derived by Boubnov *et al.* (1995) to fit their data with good agreement.

† Email address for correspondence: c.leclercq@bristol.ac.uk

However, in 2001, Molemaker *et al.* (2001) and Yavneh *et al.* (2001) discovered a new type of non-axisymmetric *primary* linear instability occurring outside the domain of axisymmetric centrifugal instability set by Rayleigh's criterion (Rayleigh 1917) of radially decreasing angular momentum (squared). The theoretical predictions were initially made in the so-called quasi-Keplerian regime of decreasing angular velocity, using the small-gap inviscid limit. Quite recently, Park & Billant (2013) extended the domain of instability to include the case of increasing angular velocity, using WKB asymptotic analysis in the limit of large axial wavenumbers (thereby relaxing the small-gap approximation of the previous authors). This last paper reached the striking conclusion that STC is in fact always linearly unstable, except for the special case of solid-body rotation. All the instabilities discovered since 2001 come in the form of non-axisymmetric modes, and correspond to resonances between boundary-trapped inertia-gravity waves. This corresponding mechanism was called 'stratorotational instability' (SRI) by Dubrulle *et al.* (2005), which assessed its potential relevance to transition to turbulence in accretion disks. The theoretical results of Molemaker *et al.* (2001) and Yavneh *et al.* (2001) in the quasi-Keplerian regime were confirmed by viscous computations (Shalybkov & Rüdiger 2005; Rüdiger & Shalybkov 2009) and experiments (Le Bars & Le Gal 2007), which also showed that non-axisymmetric instabilities could dominate within the 'centrifugally unstable domain' (as defined by Rayleigh's criterion). Le Bars & Le Gal (2007) suggested a continuous connection between these 'centrifugal modes' and SRI, which was recently established by Leclercq *et al.* (????*a*), using viscous linear stability analysis.

The potential relevance of SRI-type modes in the centrifugally unstable regime suggests to reconsider carefully the role of non-axisymmetric linear instabilities in the mechanisms of layer formation in STC, even when the outer cylinder is fixed. Using large-gap and large $Re$ (up to $10^4$) direct numerical simulations, we will show the predominant role of nonlinearly selected coherent structures formed by finite-amplitude non-axisymmetric waves in the layering process. We put forward the idea that these large vortical structures 'survive' at large $Re$ and set the depth of the well-mixed layers by overturning the density field. This idea is tested against a new series of experiments with good qualitative agreement, although much sharper interfaces in the experiments reveal the crucial role played by the Schmidt number in the homogenization process. Part 2 is dedicated to establishing a link between the coherent structures identified in this paper, and the axial buoyancy flux.

The plan of the paper is as follows. In §2 and §3, we introduce the governing equations and numerical methods. In §4, we demonstrate the dominant role played by non-axisymmetric modes in the linear dynamics of our system. In §5, we present the flow patterns obtained from direct numerical simulations, and focus on the description of the coherent structures embedded in the dynamics. We describe these structures in both spectral and physical space, in order to reveal the presence of energetic nonlinear waves, despite the large Reynolds number. In section §6, we establish a link between coherent structures and layering of the background (i.e. adiabatically sorted) density profile, highlighting the significant role played by the Schmidt number in the homogenization process. After discussing the lack of relevance of the Ozmidov length scale to this essentially non-turbulent process, we finally revisit past experiments in the light of our findings. Conclusions are presented in §7.

## 2. Governing equations

In experiments, the cylinders have a finite height $h$, and the fluid is contained by top and bottom end-plates attached to the fixed outer cylinder at $z = \pm h/2$. By contrast, we will consider infinite cylinders and assume axial periodicity for numerical simulations. The length scale $h$ will therefore represent the fundamental axial wavelength of the flow. Two nondimensional parameters describe the geometry in both cases,

$$\eta := \frac{r_i}{r_o} \quad \text{and} \quad \Gamma := \frac{h}{\Delta r}, \quad (2.1)$$

where $\Delta r := r_o - r_i$ is the gap between the outer and inner radii, $r_o$ and $r_i$. In most simulations, the radius and aspect ratios are set to $\eta = 0.417$ and $\Gamma = 3$, in order to match with the apparatus that is used for the new set of experiments described in §??. The rotation and radial shear are characterised by a Reynolds number $Re$, defined as

$$Re := \frac{r_i \Omega \Delta r}{\nu}, \quad (2.2)$$

where $\Omega$ is the rotation rate of the inner cylinder and $\nu$ is the kinematic viscosity. In most experiments, stratification is set as an initial condition on the density field, caused by controlled variations in the concentration of salt in water. The density field then freely evolves under the effects of advection and diffusion. Conservation of mass implies no-flux boundary conditions on density, eventually leading to the destruction of stratification in the long-term, whether the inner cylinder is rotated or not. This phenomenon was artificially circumvented in the numerics, where the total density field $\rho_{\text{tot}}$ was decomposed as

$$\rho_{\text{tot}} = \rho_0 + \underbrace{\bar{\rho} + \rho}_{\tilde{\rho}}, \quad (2.3)$$

where $\rho_0$ is a reference density and $\tilde{\rho}$ a deviation made of a linear background stratification of buoyancy frequency $N$ ($g$ denotes gravity),

$$\bar{\rho} = -\frac{\rho_0 N^2}{g} z, \quad (2.4)$$

and a periodic perturbation $\rho$ in both the azimuthal and axial directions. At all times, the stratification was forced through the linear term $\bar{\rho}$, and the strength of this forcing is characterised by a bulk Richardson number

$$Ri := \frac{N^2}{\Omega^2}. \tag{2.5}$$

In an experiment, this definition would only be appropriate at $t = 0$ for an initially linear stratification, but it is well-defined at all times in numerical simulations. Stratification also depends on the Schmidt number

$$Sc := \frac{\nu}{\kappa} \tag{2.6}$$

based on the diffusion coefficient of mass $\kappa$ and kinematic viscosity $\nu$. In experiments with salty water, $Sc \approx 700$, but this value is out-of-reach in the DNS, so a value of $Sc = 1$ was used instead in our simulations. Some runs at $Sc = 10$ and $Sc = 16$ were also performed in order to evaluate the impact of the Schmidt number.

The flow is governed by the incompressible Navier–Stokes equations in the Boussinesq approximation, relying on the assumption that $|\tilde{\rho}| \ll \rho_0$, and that the curvature of the isopycnals due to centrifugal effects can be neglected (see Lopez *et al.* (2013) for the Boussinesq approximation in rapidly rotating flows). In the following, we choose $\Delta r$, $r_i \Omega_i$ and

$$\Delta \rho := \bar{\rho}(z - h/2) - \bar{\rho}(z + h/2) = \frac{\rho_0 N^2 h}{g} \tag{2.7}$$

as typical length, velocity and density scales of the problem. This set of scales is used to define most nondimensional quantities throughout this article. However, there will be two exceptions: frequencies/angular velocities will be made nondimensional with respect to the inner cylinder rotation rate $\Omega$ and the unit of energy will be $\rho_0 (r_i \Omega)^2 \Delta r^3$. Note that with our choice of scales, the vertical derivative of the background stratification is $\partial_z \bar{\rho} = -1/\Gamma$, e.g. $-1/3$.

The governing equations for the velocity field $\mathbf{u}$, expressed in cylindrical coordinates as $\mathbf{u} = u\mathbf{e}_r + v\mathbf{e}_\theta + w\mathbf{e}_z$, and the perturbation density $\rho$ read

$$\partial_t \mathbf{u} + \mathbf{u} \cdot \nabla \mathbf{u} = -\nabla p + \frac{1}{Re}\nabla^2 \mathbf{u} - \beta \rho \mathbf{e}_z, \tag{2.8}$$

$$\partial_t \rho + \mathbf{u} \cdot \nabla \rho - \frac{w}{\Gamma} = \frac{1}{Re\, Sc}\nabla^2 \rho, \tag{2.9}$$

$$\nabla \cdot \mathbf{u} = 0, \tag{2.10}$$

where $p$ is a potential based on the actual pressure and finally

$$\beta := \Gamma \frac{(1-\eta)^2}{\eta^2} Ri \tag{2.11}$$

is the nondimensional version of the reduced gravity $g\Delta\rho/\rho_0$. As explained in Leclercq *et al.* (2016), 'centrifugal buoyancy' needs not be added to our model since the Reynolds number is always below $10^4$. Given the choice of velocity scale, the boundary conditions on the velocity field are

$$\mathbf{u} = \begin{cases} (0, 1, 0) & \text{at the inner cylinder } r_i = \eta/(1-\eta), \\ (0, 0, 0) & \text{at the outer cylinder } r_o = 1/(1-\eta). \end{cases} \tag{2.12}$$

The no-flux boundary conditions on the perturbation density field correspond to a vanishing radial derivative $\partial_r \rho = 0$.

## 3. Numerical methods

Given the azimuthal and axial periodicities, all fields $\mathbf{q} := (\mathbf{u}, p, \rho)$ can be decomposed into Fourier modes in $\theta$ and $z$:

$$\mathbf{q}(r,\theta,z,t) = \sum_{m=-n_\theta/2}^{n_\theta/2} \sum_{k=-n_z/2}^{n_z/2} \mathbf{q}_{m,k}(r,t) \exp[\mathrm{i}(m\theta + kk_0 z)], \qquad (3.1)$$

where $k_0 := 2\pi/\Gamma$ and $\mathbf{q}_{-m,-k} = \mathbf{q}_{m,k}^*$ (* denotes the complex conjugate). A pseudospectral method was implemented to time-march the coefficients $\mathbf{q}_{m,k}$, using the code of Shi *et al.* (2015). In the radial direction, high-order finite differences were used, using 9-point stencils except in the direct vicinity of the walls. A distribution of Gauss–Lobatto points was used in order to resolve boundary layers efficiently. The diffusive terms were treated implicitly using backward differentiation while the nonlinear terms were extrapolated with a second-order Adams–Bashforth method. The time-step was fixed for each simulation such as to ensure convergence (Shi *et al.* (2015) indicate that spatial discretization is the main source of error in the code, not temporal discretization). Stratification had been added to the code prior to our study, but with Dirichlet boundary conditions for the density at the walls and no vertical forcing in the form of (2.4). Straightforward modifications allowed us to take into account the specificity of the present problem. The modifications have been validated by checking the growth rates of linear instabilities against theoretical predictions (see figure 13(*b*) for instance). Details on the linear stability solver are reported in (Leclercq *et al.* ????*a*).

Simulations were started with an incompressible perturbation of kinetic energy equal to $10^{-6}$ times the kinetic energy of the base flow, by forcing equally each mode in the range $-n_z/4 \leqslant k \leqslant n_z/4$ and $-n_\theta/4 \leqslant m \leqslant n_\theta/4$ (only the vertical and azimuthal components of the velocity field were perturbed). The phase of the modes were either chosen randomly or synchronized, sometimes leading to different behaviours (more details on the effect of initial conditions in §5.3 and §6.4). The steady boundary conditions (2.12) were reached after a linear spin-up of the inner cylinder, over a time scale $\tau$ of 2 s in real time, in order to model an impulsive start, or 1 h to model a slow ramp (for the $\eta = 0.417$ apparatus of §?? with $\Delta r = 140$ mm). Simulations typically lasted for one thousand convective time units $d/(r_i \Omega)$, in order to capture a statistically steady state. Transient motions typically last $\approx 400$ convective time units following an impulsive start (around $\approx 100$ rotations of the inner cylinder for a $\eta = 0.417$ apparatus). Table 1 summarizes the physical and numerical parameters for all simulations, together with a list of output quantities which will be defined and commented throughout the rest of this paper.

Spatial convergence was checked by ensuring that the absolute value of the trailing spectral coefficients of the radial velocity field (after discrete Chebyshev transform of $u_{m,k}(r,t)$ at fixed $t$) in all directions were always at least four orders of magnitude lower than the maximum coefficient.

## 4. Linear analysis

In the spectral decomposition (3.1), each Fourier mode is denoted by a pair of integer wavenumbers $(m,k)$: $m$ in the azimuthal direction and $k$ in the axial. Modes with $m = 0$ and $k \neq 0$ correspond to toroidal structures, $m \neq 0$ and $k = 0$ are axially invariant, and finally $mk \neq 0$ have a helical shape. If $mk < 0$, the helix is said to be 'right-handed', otherwise it is 'left-handed'. Linear analysis entails studying independently the properties of each spatial mode of an infinitesimal perturbation $\mathbf{q}' := \mathbf{q} - \mathbf{Q}$, as it interacts with

| | A1 | A2 | B | C | D | E | F | G | S | CT |
|---|---|---|---|---|---|---|---|---|---|---|
| $Re$ | 2000 | | 5000 | 5000 | 5000 | 10000 | 5000 | 5000 | 245 | 816 |
| $Ri$ | 3 | | 2 | 3 | 10 | 3 | 3 | 3 | 4.15 | 0.373 |
| $Sc$ | 1 | | 1 | 1 | 1 | 1 | 10 | 1 | 16 | 16 |
| $\eta$ | 0.417 | | 0.417 | 0.417 | 0.417 | 0.417 | 0.417 | 0.625 | 0.769 | 0.769 |
| $\Gamma$ | 3 | | 3 | 3 | 3 | 3 | 3/8 | 3 | 3.25 | 3.25 |
| $n_r$ | 64 | | 80 | 80 | 80 | 112 | 192 | 96 | 96 | 96 |
| $n_\theta$ | 126 | | 256 | 256 | 256 | 448 | 720 | 384 | 96 | 192 |
| $n_z$ | 288 | | 400 | 400 | 400 | 672 | 192 | 576 | 128 | 192 |
| $\Delta t$ | $2 \times 10^{-3}$ | | $10^{-3}$ | $10^{-3}$ | $10^{-3}$ | $7 \times 10^{-4}$ | $3 \times 10^{-4}$ | $10^{-3}$ | 0.02 | $2.5 \times 10^{-3}$ |
| $m_1, m_2$ | 2, 2 | 2, 1 | 1, 2 | 1, 2 | 2, 2 | 1, 1 | 1, 2 | 2, 2 | 3, 3 | 0, 0 |
| $k_1, k_2$ | 6, −6 | 6, −7 | ±7, −7 | −8, 8 | 9, −9 | 9, −10 | −1, 1 | 6, −6 | 3, −3 | ±1, ±2 |
| $\omega_1, \omega_2$ | 0.48, 0.48 | 0.47, 0.21 | 0.19, 0.38 | 0.20, 0.41 | 0.60, 0.60 | 0.18, 0.18 | 0.18, 0.38 | 0.57, 0.57 | 1.4, 1.4 | — |
| $k_\text{axi}$ | 16 | 20 | 21 | 24 | 18 | 19 | 3 | 12 | 6 | — |
| $\omega^\star$ | 0.24 | 0.22 | 0.19 | 0.20 | 0.30 | 0.18 | 0.19 | 0.29 | 0.47 | — |
| $\langle v/r \rangle_{V,t}$ | 0.19 | 0.19 | 0.17 | 0.18 | 0.25 | 0.17 | 0.16 | 0.28 | 0.40 | 0.36 |
| $Re_b$ | 1.3 | 1.2 | 3.5 | 2.2 | 0.58 | 3.6 | 2.8 | 12 | 5.4 | 123 |
| $l_O \, (\times 10^{-2})$ | 1.6 | 1.6 | 1.9 | 1.4 | 0.51 | 1.2 | 1.5 | 4.9 | 19 | 91 |
| $\sigma^{1/2} \, (\times 10^{-2})$ | 0.98 | 0.91 | 1.2 | 0.93 | 0.54 | 0.90 | 1.1 | 1.9 | 3.1 | 16 |

TABLE 1. Summary of simulation parameters (top part of the table) and output metrics related to layer formation (bottom part) and defined throughout this paper. A1 and A2 correspond to two different initial conditions, leading respectively to a ribbon and a mixed-ribbon state. Notice that F was carried out in a box which is 8 times smaller in the axial direction than other simulations. By extrapolating to the full box $\Gamma = 3$, we would recover the same values of $k_1, k_2, k_\text{axi}$ as in run C. The right side of the table (runs S and CT) will be discussed in §6.3. For CT, the flow is not simply dominated by a pair of helical modes, even though there are some powerful helical structures in the flow. Therefore, we do not provide values for $k_\text{axi}$ or frequencies.

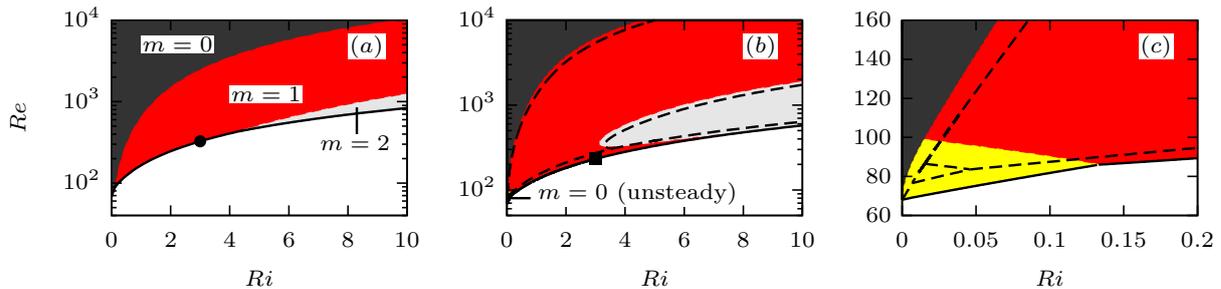

FIGURE 1. Azimuthal wavenumber $m$ associated with fastest growing mode as a function of $Re$ and $Ri$ for (a) $Sc = 1$, (b) $Sc = 7$ (dashed lines) and $Sc = 700$. The dominance region of the unsteady $m = 0$ mode for $Sc = 7, 700$ is visible in (c), which is a zoom of (b) at low $Ri$ and $Re$. Symbols (solid dot in (a) and solid square in (b)) indicate the parameter values for the computation of the critical modes shown in $2(a, b)$.

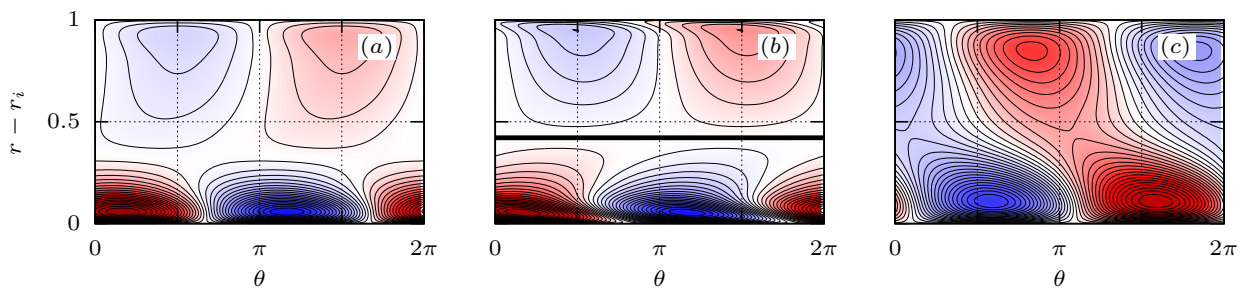

FIGURE 2. Vertical velocity component of critical eigenmode for $Ri = 3$: (a) $\eta = 0.417$, $Sc = 1$, $Re = 325.9$ (solid dot in $1(a)$), (b) $\eta = 0.417$, $Sc = 700$, $Re = 231.7$ (solid square in $1(b)$), (c) $\eta = 0.625$, $Sc = 1$. Red/blue regions have opposite sign. The thick solid line in (b) indicates the location of a critical layer $\omega_r = \Omega(r)$.

the base flow $\mathbf{Q}$ of velocity

$$\mathbf{U} = \left(0, \frac{\eta}{1+\eta}\left[-r + \frac{1}{(1-\eta)^2 r}\right], 0\right) \tag{4.1}$$

and density $\bar{\rho}$. The base flow being steady, the time-dependence of each disturbance mode is of the form

$$\mathbf{q}'_{m,k}(r,t) = \hat{\mathbf{q}}'_{m,k}(r)\exp(-i\omega t), \tag{4.2}$$

where $\omega = \omega_r + i\omega_i$ is a complex frequency. The temporal growth rate corresponds to $\omega_i$ (positive/negative for growth/decay) and $\omega_r$ is the frequency. The symmetries of the linearized Navier–Stokes–Boussinesq equations (by complex conjugation and by reflection $z \to -z$) are such that

$$\omega(m, k) = \omega(m, -k) = -\omega^*(-m, k) = -\omega^*(-m, -k). \tag{4.3}$$

This means that $m$ can be assumed positive without loss of generality, and that for a given $m$, left- and right-helices of same $|k|$ have identical growth rate and frequency. But waves propagating against the base flow rotation $\omega_r < 0$ have a different growth rate from those propagating with the base flow $\omega_r > 0$.

Figure 1 shows the azimuthal wavenumber of the fastest growing mode, after optimization of the temporal growth rate over all axial wavenumbers $k$ (in this section $k$ is assumed real). We find that non-axisymmetric modes usually dominate, except in two cases: large $Re$ (i.e. far from instability threshold, where linear theory no longer applies) or both small $Ri$ and $Re$ when the $Sc$ number is large enough. This finding is at odds with the assumption sometimes found in the literature (Le Dizès & Riedinger 2010;

Oglethorpe *et al.* 2013) that axisymmetric modes dominate the linear dynamics when the outer cylinder is fixed. As $Ri$ increases, the critical mode switches from $m = 1$ to $m = 2$.

The main effect of the $Sc$ number is the appearance of a tiny zone of dominance of an unsteady axisymmetric mode at low $Ri$ and $Re$, this zone being absent from the $Sc = 1$ diagram. Otherwise, changing the Schmidt number by a factor of 100 in figure 1(*b*) leads to qualitatively similar pictures at $Sc = 7$ and $Sc = 700$, despite a weak stabilizing effect of $Sc$. These results are consistent with the (axisymmetric) computations of Hua *et al.* (1997) for $\eta = 0.776$, who found virtually no effect of the $Sc$ number on the value of the critical Reynolds number beyond $Sc \approx 16$. There are however some differences between the stability properties at $\eta = 0.417$ and $\eta = 0.769$, which will be discussed in **??**. The main conclusion of linear analysis at $\eta = 0.417$ is that non-axisymmetric modes bifurcate first at finite $Ri$.

Figure 2 represents the critical eigenmode found for $Ri = 3$. We show the effect of the Schmidt number and the radius ratio. We first notice that the structure of the critical mode is virtually independent of $Sc$. But whereas the modes are essentially localized at the inner boundary for $\eta = 0.417$, there is also a significant contribution at the outer boundary for the smaller-gap case $\eta = 0.625$.

All three figures are reminiscent of figure 2(*b*) in Park & Billant (2013), showing a typical stratorotational instability (SRI) mode: there are waves localized at each boundary, out-of-phase in $\theta$ by approximately $\pi/2$ (although in that study, the outer cylinder rotates and the inner one is fixed). At large $Sc$, we even find a critical layer (azimuthal phase speed of the mode locally equal to the angular velocity of the base flow) between the two boundary-trapped waves, as expected from the WKB analysis of Park & Billant (2013). The structure of the modes is also consistent with the characteristics of SRI initially depicted by Yavneh *et al.* (2001); Molemaker *et al.* (2001) in the quasi-Keplerian regime.

However, despite all these similarities with SRI modes, we note that it is impossible in practice to assign a specific instability mechanism to non-axisymmetric modes in the centrifugally unstable regime. Indeed, at finite $Re$, the two distinct instability mechanisms couple and may therefore not be distinguished from one another (Leclercq *et al.* ????*a*). Regardless, non-axisymmetric modes bifurcate first in the presence of stratification in the $\eta = 0.417$ apparatus.

## 5. Coherent structures: (mixed)-ribbons and (mixed)-cross-spirals

In this section, we provide a description of the patterns obtained in the fully nonlinear regime, when individual perturbation modes interact with each other and not only with the base flow.

### 5.1. *In spectral space*

Figure 3(*a*, *b*) shows the dominant Fourier modes of the perturbation velocity $\mathbf{u}'$ in terms of kinetic energy

$$E_{m,k}^{k'} := \pi \Gamma \int (|u'_{m,k}|^2 + |v'_{m,k}|^2 + |w'_{m,k}|^2) r \, \mathrm{d}r \tag{5.1}$$

of the statistically steady final states reached in simulations C and D. In both cases, the dynamics is clearly dominated by a pair of modes (and their complex conjugates, not shown in the figures). Each pair consists of two helical modes of opposite handedness (earlier defined as the sign of $mk$) and comparable level of kinetic energy. The next energetic modes are clearly generated by interactions of the primary modes and their complex conjugate. Moreover, the spatiotemporal spectra in figure ??(c, d) clearly indicate that to each spatial mode corresponds a well-defined frequency: this is the signature of nonlinear waves. To summarize, energetic modes belong to a vector space in $(m, k, \omega)$ generated by the two dominant modes. This type of coherent structure is called *ribbon* (Demay & Iooss 1984) in case (a) or *mixed-ribbon* (Altmeyer & Hoffman 2014) in case (b), depending whether the two helices are images of one another or not:

$$\text{(Pure-)ribbon in run D:} \quad (m, k, \omega) \in \text{Span}\{(1, 9, \omega_{1,9}), (1, -9, \omega_{1,-9})\} \tag{5.2}$$

$$\text{Mixed-ribbon in run C:} \quad (m, k, \omega) \in \text{Span}\{(2, -8, \omega_{2,-8}), (1, 8, \omega_{1,8})\} \tag{5.3}$$

The axisymmetric subspace of these structures has a fundamental axial wavenumber $k_\text{axi}$ and a frequency $\omega_\text{axi}$ given by

$$(0, k_\text{axi}, \omega_\text{axi}) = \delta(m_1, k_1, \omega_{m_1, k_1}) + \gamma(m_2, k_2, \omega_{m_1, k_1}), \tag{5.4}$$

where $(\delta, \gamma)$ is a couple of integers such that the azimuthal wave number on the left-hand side is zero and that $k_\text{axi}$ is positive and minimal. For case C, $(\delta, \gamma) = (1, -2)$, leading to $k_\text{axi} = 24$ axisymmetric vortices drifting vertically at a phase speed $\omega_\text{axi}/(k_\text{axi} k_0) \neq 0$ because $\omega_{1,8} \neq \omega_{2,-8}/2$. However, these two dominant frequencies, $\omega_{1,8}$ on the one hand and $\omega_{2,-8}/2$ on the other, have very close values (see figure 3c), leading to slow oscillations at $\omega_\text{axi} \ll 1$. For the pure-ribbon D though, $(\delta, \gamma) = (1, -1)$, leading to $(k_\text{axi}, \omega_\text{axi}) = (18, 0)$, i.e. 18 axisymmetric and stationary Taylor vortices.

In the remainder of this paper, we will drop the frequency $\omega_{m,k}$ and denote these structures as $\text{Span}\{(m_1, k_1), (m_2, k_2)\}$ to simplify notations. Finally, (mixed)-cross-spirals correspond to branches where one of the helices is more energetic than the other (Altmeyer & Hoffman 2014).

Ribbon structures were initially found in counter-rotating unstratified Taylor–Couette flow, where non-axisymmetric modes are also able to dominate the linear dynamics (Demay & Iooss 1984; Langford *et al.* 1988; Tagg *et al.* 1989). Purely helical and ribbon branches bifurcate simultaneously, but Pinter *et al.* (2006) showed that ribbons are unstable close to the bifurcation point. The stability transfer from the helical branch to the ribbon branch is mediated by cross-spirals. Altmeyer & Hoffmann (2010) later showed that mixed-cross-spirals (helices of different pitch $m/(kk_0)$ in absolute value and opposite handedness $\text{sgn}(mk)$) also bifurcate out of saturated helical branches. A mixed-cross-spiral can either branch back to the helical structure from which it was created ('bypass' scenario; Altmeyer & Hoffmann (2010)), or connect to the helical branch of opposite handedness, generating a 'footbridge' (Altmeyer & Hoffman 2014) in the bifurcation diagram. Along this footbridge, the ratio of amplitude between the two dominant helical modes vary from 0 to infinity. When this ratio is equal to one, a mixed-ribbon is formed.

The computations of Pinter *et al.* (2006); Altmeyer & Hoffmann (2010); Altmeyer & Hoffman (2014) were carried out at Reynolds numbers $O(10^2)$ (one for each cylinder), but here we observe the signature of the nonlinear branches far from the instability threshold. The flow is therefore strongly nonlinear, as can be seen from the large mean flow distortion $E_{0,0}^{k'}$. It is remarkable that these coherent structures 'survive' at such high Reynolds numbers, at least one order of magnitude larger than the instability threshold, and after an impulsive start. Just like turbulent Taylor vortices in unstratified Taylor–Couette flow at similar $Re$ (Brauckmann & Eckhardt 2013), they correspond to strongly attracting saddles in phase space. Both spatial and spatiotemporal spectra are not always as sharply peaked as in figure 3, but there is always the signature of a powerful coherent structure in the flow. For low enough $Re$ and $Sc$ or large enough $Ri$, the flow is laminar and the coherent structures are true nonlinear attractors (simulations A1, A2 and D, see next section).

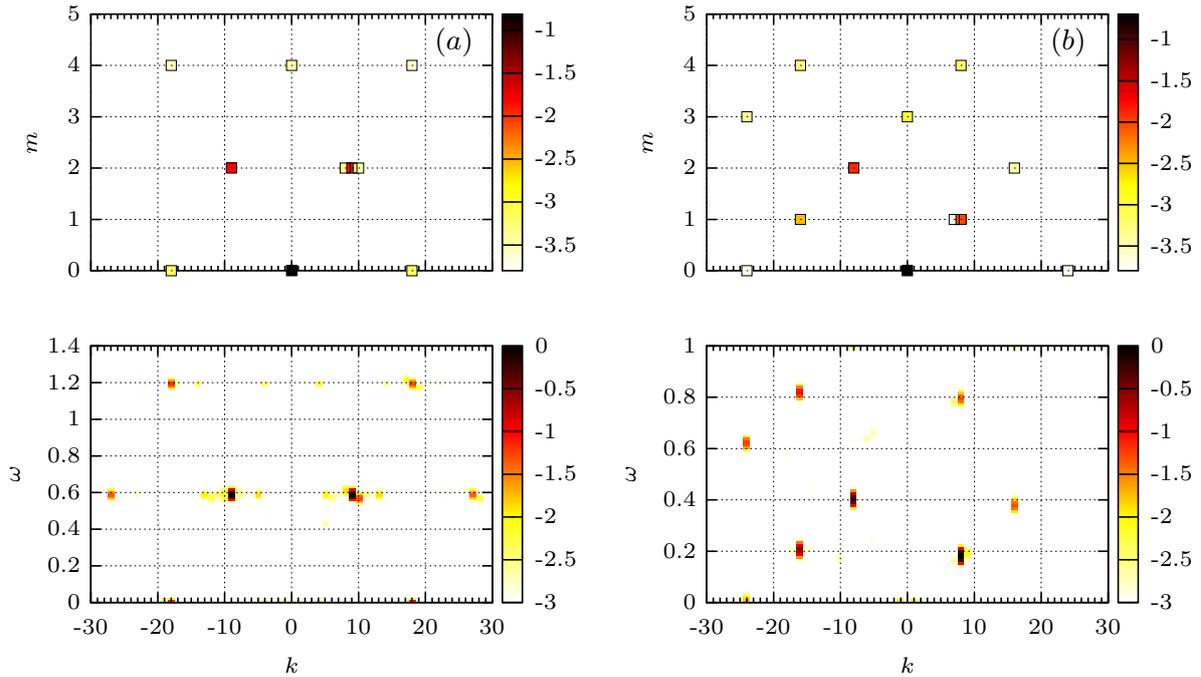

FIGURE 3. $(a,b)$ Time-averaged spatial spectrum of perturbation kinetic energy (normalized by kinetic energy of the base flow). $(c,d)$ Power spectral density (psd) of $\langle \rho(r,\theta,z,t)\rangle_r$ (normalized by the maximum value) for a fixed angle $\theta$, from two-dimensional Fourier transform of spatiotemporal diagrams (with Hanning windowing and oversampling). The unit of $\omega$ is $\Omega$. Only the most energetic modes are plotted. To take into account complex conjugate modes, which have the same modal energy (resp. psd), we plot $E_{m,k}^{k'} + E_{-m,-k}^{k'} = 2\,E_{m,k}^{k'}$ (resp. $\mathrm{psd}_{m,k} + \mathrm{psd}_{-m,-k} = 2\,\mathrm{psd}_{m,k}$) when $m \neq 0$, and just $E_{0,k}^{k'}$ (resp. $\mathrm{psd}_{0,k}$) otherwise. $(a,c)$ Run D, $(b,d)$ run C.

### 5.2. *In physical space*

#### 5.2.1. *Snapshots*

The flow patterns generated by these interactions of modes are represented in this section, with slices of the vertical density derivative at constant radius $r - r_i = 10\%$ in figure 4 and at constant angle $\theta$ in figure 5. All coherent structures form sharp interfaces in the stratified fluid. The interfaces are inclined in the $(\theta, z)$-plane when a saturated helical branch or cross-spiral locally dominate (for instance case G), slightly tilted for mixed-ribbons (A2, C and F), and horizontal for a 'pure' ribbon (A1 and D). The tilt angle for cross-spirals and mixed-ribbons is caused by the broken symmetry between interacting helices. The sharpness of the interfaces depends on the different parameters: it increases with $Re$, $\eta$ and $Sc$, but decreases with $Ri$. In other terms, the more disorganized the flow, the sharper the interfaces.

Between these sharp interfaces, well-mixed layers form, with vertical density derivative $\partial_z \tilde{\rho}$ close to zero. We note however that these homogeneous layers seem localized in the annulus. They are clearly not axisymmetric as represented in figure 1 of Oglethorpe *et al.* (2013). Moreover, they also seem to be localised radially, as can be seen in figure 4. The radial localisation seems consistent with the structure of the linear critical modes in figure 2: the wave trapped at the outer cylinder has a large amplitude for $\eta = 0.625$, resulting in a sharper interface than for $\eta = 0.417$.

Superimposing the meridional velocity field to the vertical density derivative contours reveals the presence of radial jets at the location of interfaces (these were already noted by Hua *et al.* (1997)). Sharpening of the density profile therefore coincides with the large-scale structures of the flow, indicating in turn that layer depth may be governed by the same nonlinear mechanism which selects these vortical structures in the first place.

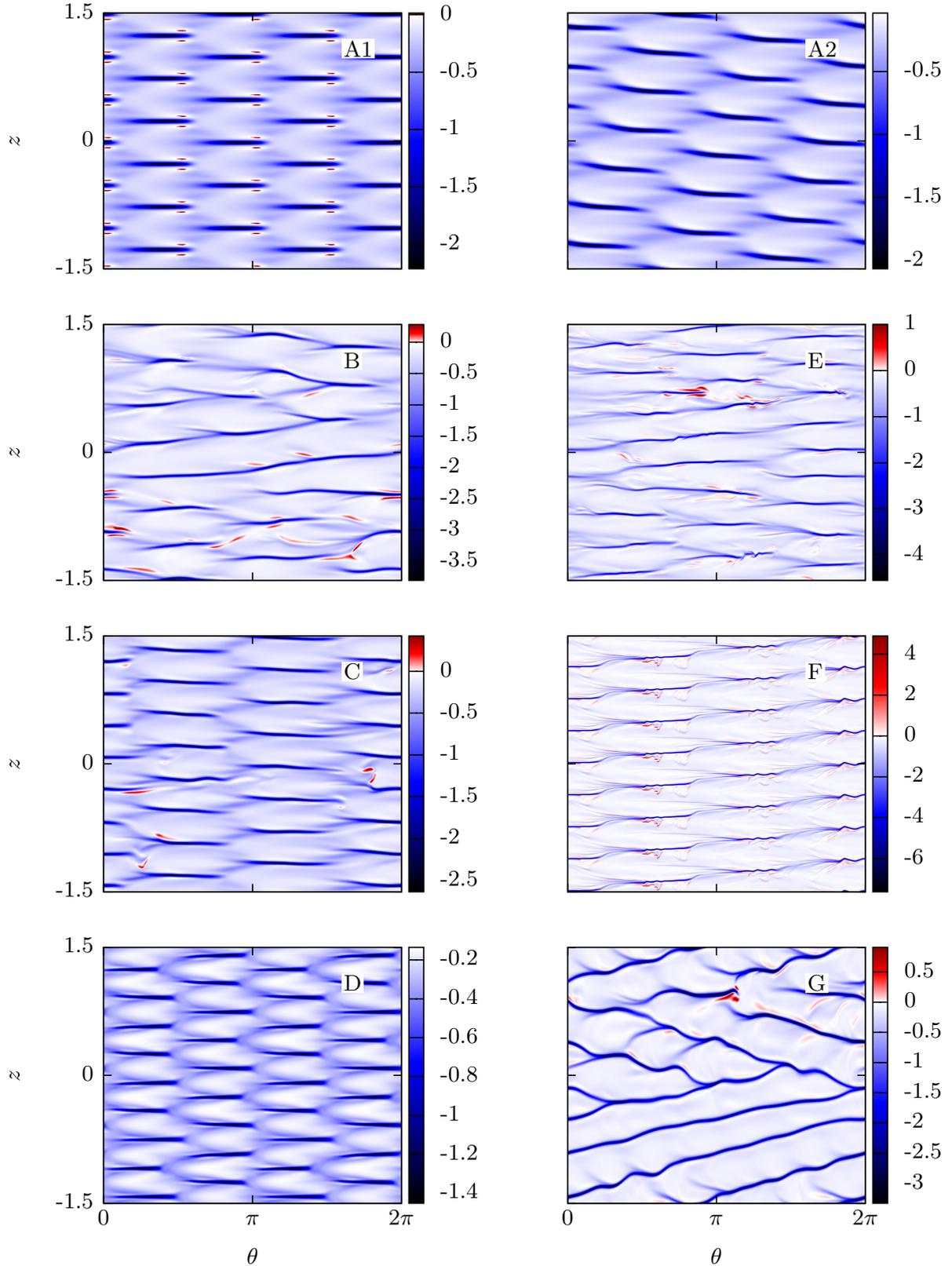

FIGURE 4. Snapshots of vertical derivative of density field $\partial_z \tilde{\rho}$ at fixed $r - r_i = 10\%$ for runs in table 1. The figure labels match the name of the simulations. Simulation F has been performed over a vertical extent which is 8 times smaller than for other simulations ($\Gamma = 3/8$ instead of 3). In this representation, we have artificially extended the computational domain to $\Gamma = 3$ (using axial periodicity) and rescaled the density field accordingly for comparison.

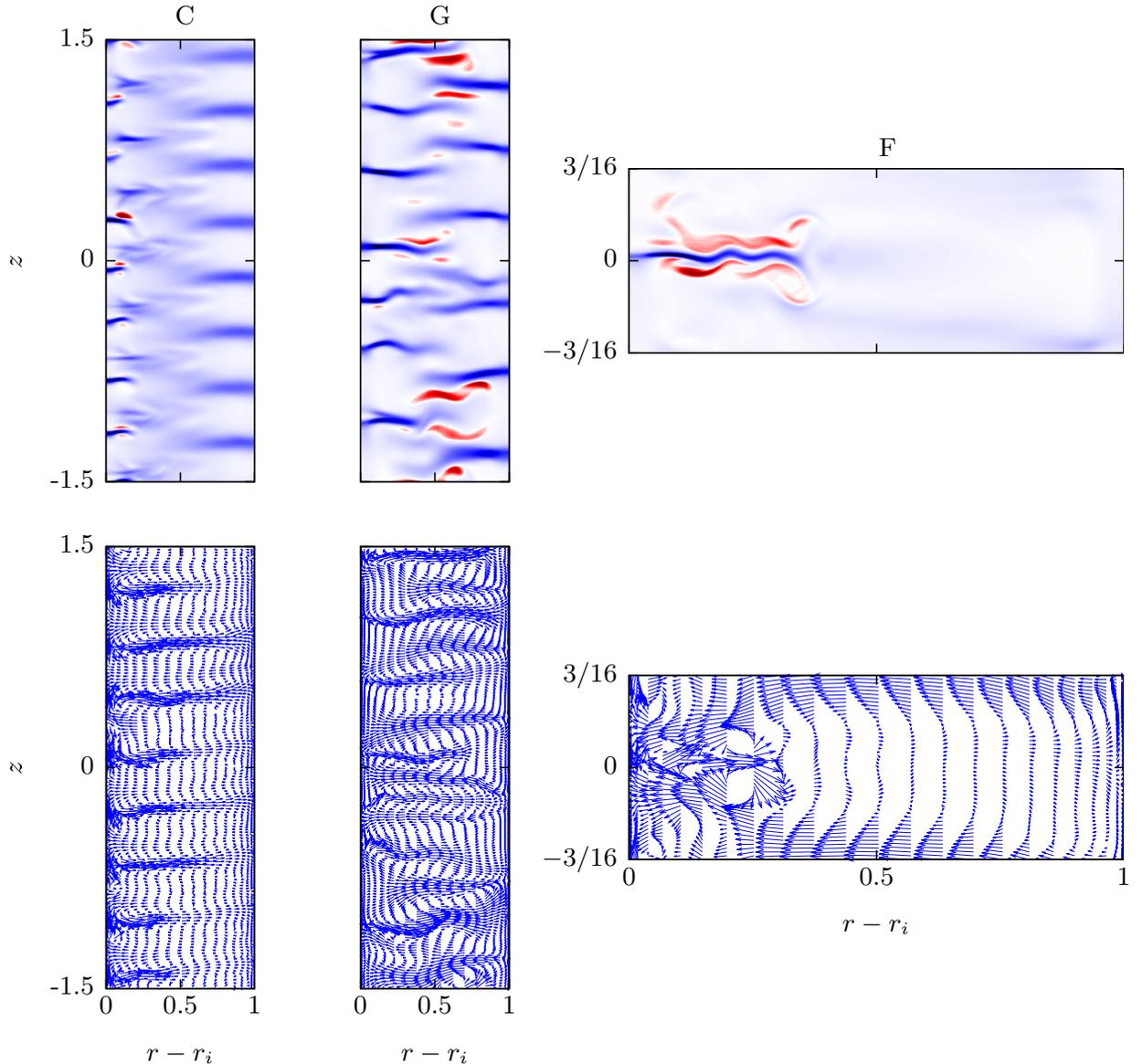

FIGURE 5. Snapshots of vertical derivative of density field $\partial_z \tilde{\rho}$ at fixed $\theta$. The meridional velocity field is shown on (a) and (c), but not on (b) for clarity. Runs (a) C, (b) F, (c) G. As in figure 4, run F is shown over a domain which is 8 times bigger than the actual numerical 'box'. The density scale $\Delta \rho$ has been changed accordingly.

Finally, figure 6 shows three-dimensional representations of (a) the radial velocity component $u$ and (b) perturbation density $\rho$, for simulation F at $Sc = 10$. There is clear evidence of turbulent structures in both representations. The 3 'lobes' in (a) are the signature of the axially invariant mode $(3,0) = (2,-1) + (1,1)$ emerging from the interaction of the dominant helices $(2,-1)$ and $(1,1)$. The radial jet from the inner cylinder, visible in (a), coincides with the sharp density front visible in (b). From this top view, the front takes the form of a spiral arm. Because the coherent structure is a mixed-ribbon, the three radial jets occur at different heights in 6(a) (see also figure 4(g)), therefore only one spiral arm is visible in the slice at constant $z$ of figure 6(b).

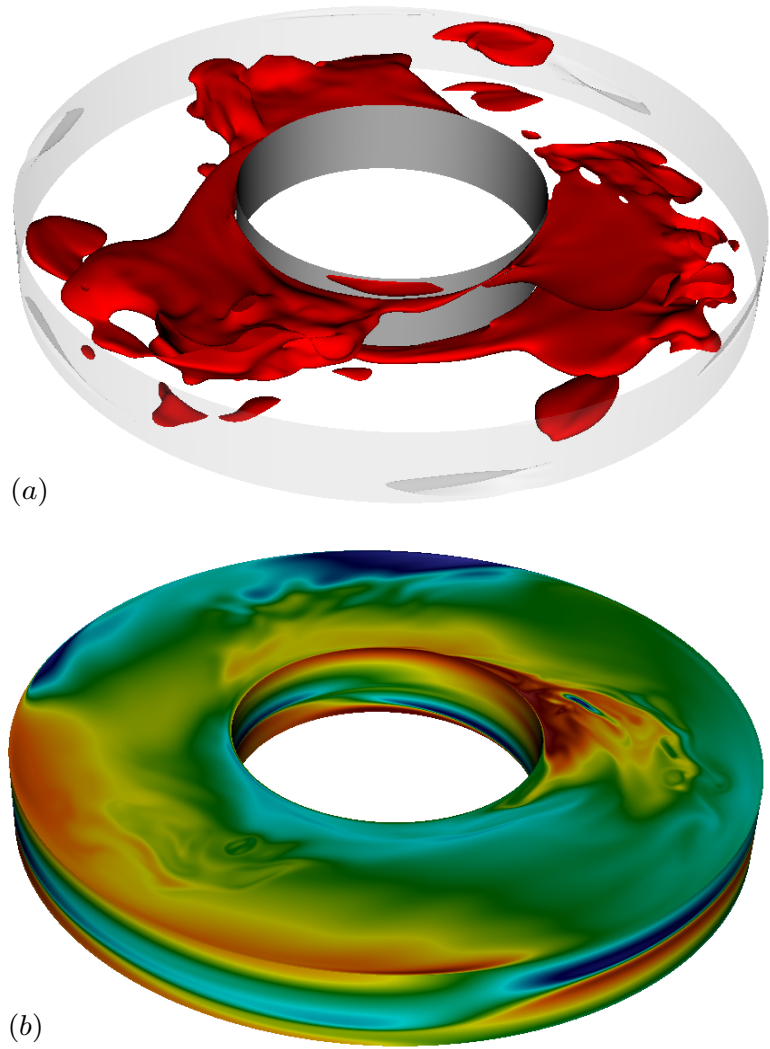

FIGURE 6. Three-dimensional visualization of run F: (*a*) isosurface of $u = 0.04$ at the end of the simulation, (*b*) perturbation density field $\rho$ at the same time. The inner cylinder is rotating counterclockwise.

5.2.2. *Spatiotemporal diagrams*

Introducing the mean azimuthal phase speed

$$\omega^\star = \frac{1}{2}\left(\frac{\omega_1}{m_1} + \frac{\omega_2}{m_2}\right), \quad (5.5)$$

we see that all helical waves rotate at about $\omega^\star \approx 0.2 - 0.3$ the angular velocity of the inner cylinder. This slow value is well approximated by the mean angular velocity of the flow $\langle v/r\rangle_{\mathcal{V},t}$, as can be seen in table 1, with little dependence on the different control parameters otherwise. This observation is consistent with the experimental findings of Le Bars & Le Gal (2007) for the stratorotational instability.

Helical waves also propagate vertically, but in the case of a pure ribbon, the symmetry between interacting modes yields a standing wave in the vertical direction. This leads to an axially frozen pattern which only rotates azimuthally, giving the impression of 'flashing' horizontal interfaces, as can be seen in the spatiotemporal diagrams of run D, figure 7(*a, c*). The interfaces are long-lived because of the relatively slow angular velocity of the pattern. We stress that this periodic appearance and disappearance of horizontal interfaces at a given height $z$ is not a secondary instability of a horizontal interface initially created by an axisymmetric Taylor vortex: it is the manifestation of a rotating pattern caused by the interaction between two symmetric helical vortices.

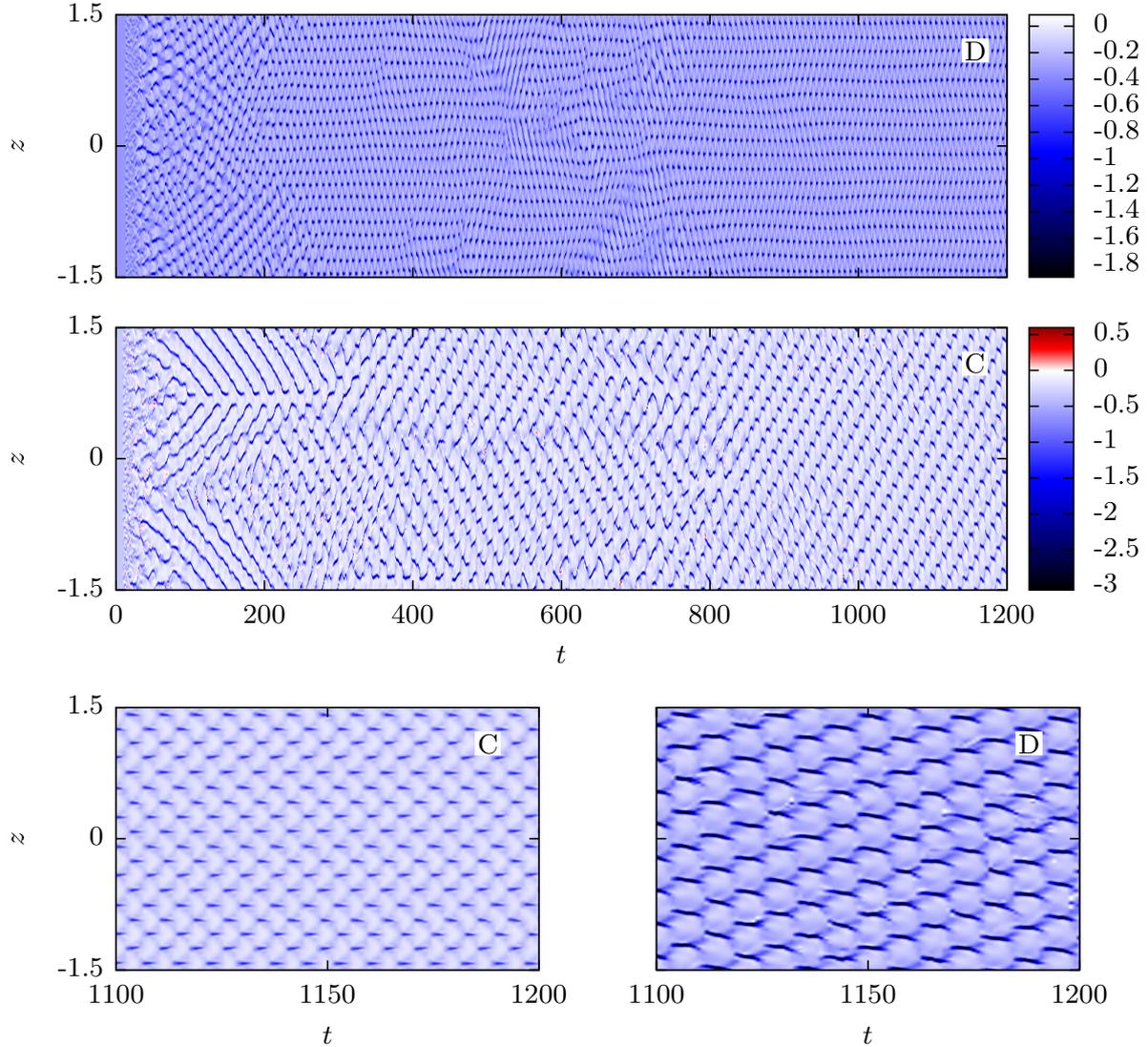

FIGURE 7. $(a, b)$ Spatiotemporal diagrams of $\partial_z \tilde{\rho}(r, \theta, z, t)$ at fixed $\theta$ and $r - r_i = 10\%$ and $(c, d)$ zooms on seemingly 'flashing interfaces'. $(a, c)$ run D (ribbon end-state), $(b, d)$ run C (mixed-ribbon end-state).

In the case of a mixed-ribbon, there may be a slow vertical drift of the interfaces caused by the difference in axial phase speed between the two interacting helices, as already explained in the previous section. This slow drift is visible in the spatiotemporal diagram for run C in figure 7(*b*,*d*). Note that the drift may be directed upwards or downwards with equal probabilities.

### 5.2.3. *Mean radial profiles*

Mean radial profiles are obtained by averaging over the two periodic directions and time. Figure 8 gives a representation for the larger $Re = 10000$ case (simulation E). The angular momentum (figure 8(*a*)) profile strongly deviates from the laminar solution, indicating strong nonlinearity of the flow. There are boundary layers at both cylinders, and the mean angular momentum slowly increases in the core. This profile is quite similar to the one obtained for turbulent Taylor vortex flow in the unstratified case (see figure 10(*b*) in Brauckmann & Eckhardt (2013)). The rms of the radial velocity (figure 8(*b*)) is dominated by the contribution of the radial jets coming off the inner cylinder (see

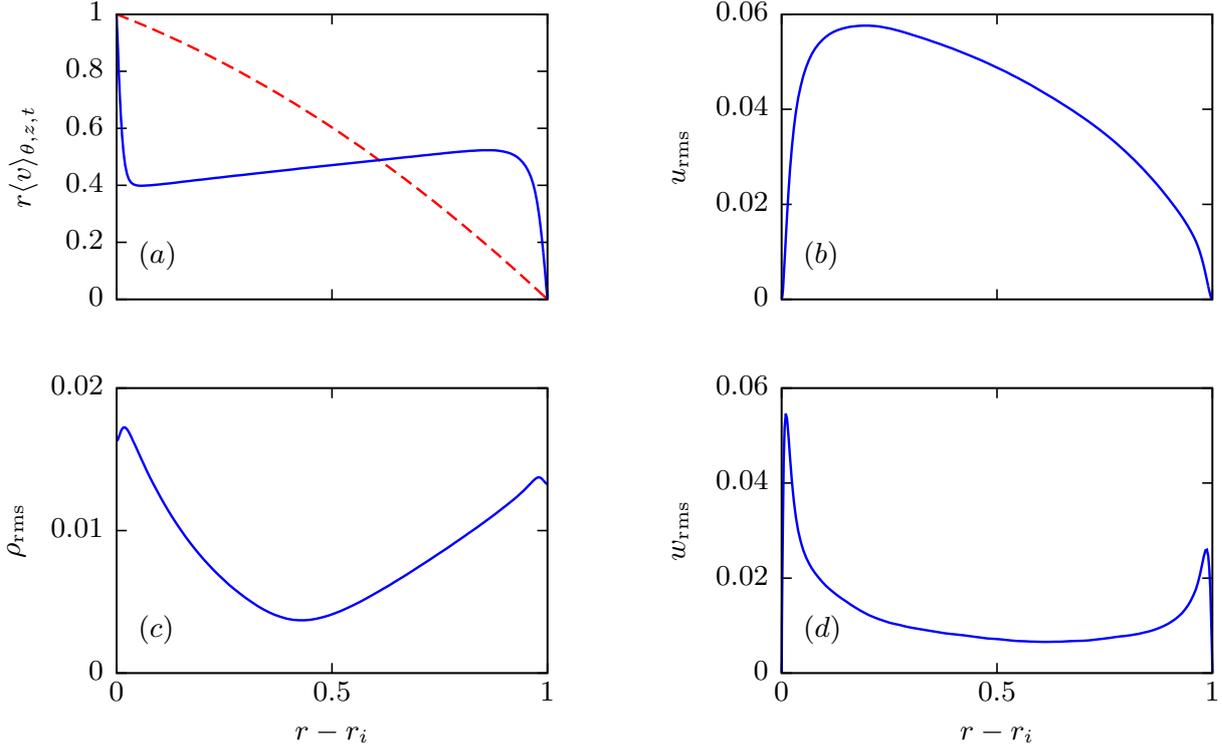

FIGURE 8. Radial profiles of flow quantities averaged with respect to time and the two periodic directions, for simulation E: (*a*) angular momentum (solid line: simulation E, dashed line: laminar case), normalized by $r_i(r_i \Omega)$. Root-mean square of (*b*) radial velocity $u$, (*c*) vertical velocity $w$ and (*d*) density perturbation $\rho$.

figure 5): the maximum radial velocity is located here around $r - r_i = 20\%$ in this case. These radial jets are responsible for the homogenization of angular momentum observed in figure 8(*a*). The fluctuations of vertical velocity and density (8(*c*, *d*)) are both localised near the walls, which is consistent with the structure of the critical linear modes in figure 2. Finally, we note the low levels of meridional flow perturbations, $< 6\%$ of the inner cylinder velocity, and the even lower rms of the perturbation density, which is only $< 2\%$ of the density difference between the bottom and top ends of the domain.

### 5.3. *Effects of initial conditions*

We tried to use linear analysis to predict the axial wavenumber of the dominant helical modes. This approach proved ineffective as the simulations are carried out too far from the primary instability threshold. However, applying linear analysis on a transiently evolving base flow yielded excellent results for estimating the dominant mode in the early stages of a simulation. The analysis was carried out by assuming the base flow to be 'frozen' at each time, as in Kim *et al.* (2004). Two types of initial conditions were considered: impulsive start ($\tau < 1$ or $2\,\mathrm{s}$ lab time) and slow linear ramp ($\tau \approx 918$ or $1\,\mathrm{h}$ lab time). In both cases, the critical axial wavenumber shoots off transiently, before decreasing back to an asymptotic value corresponding to the steady base flow (see figure 9 for $\tau = 1\,\mathrm{h}$ lab time). As a consequence, small scales structures develop transiently, before the flow eventually relaxes to an unrelated attractor (or saddle).

This relaxation can occur in the form of Eckhaus-type bifurcations (Tuckerman & Barkley 1990), as can be seen for run D in figure 7(*a*): the flow transiently latches onto mixed-ribbon $\mathrm{Span}\{(2, 11), (2, -10)\}$ for $270 \leqslant t \leqslant 360$ before suddenly jumping to other branches, until the flow finally settles down onto ribbon $\mathrm{Span}\{(2, 9), (2, -9)\}$ after $t = 800$.

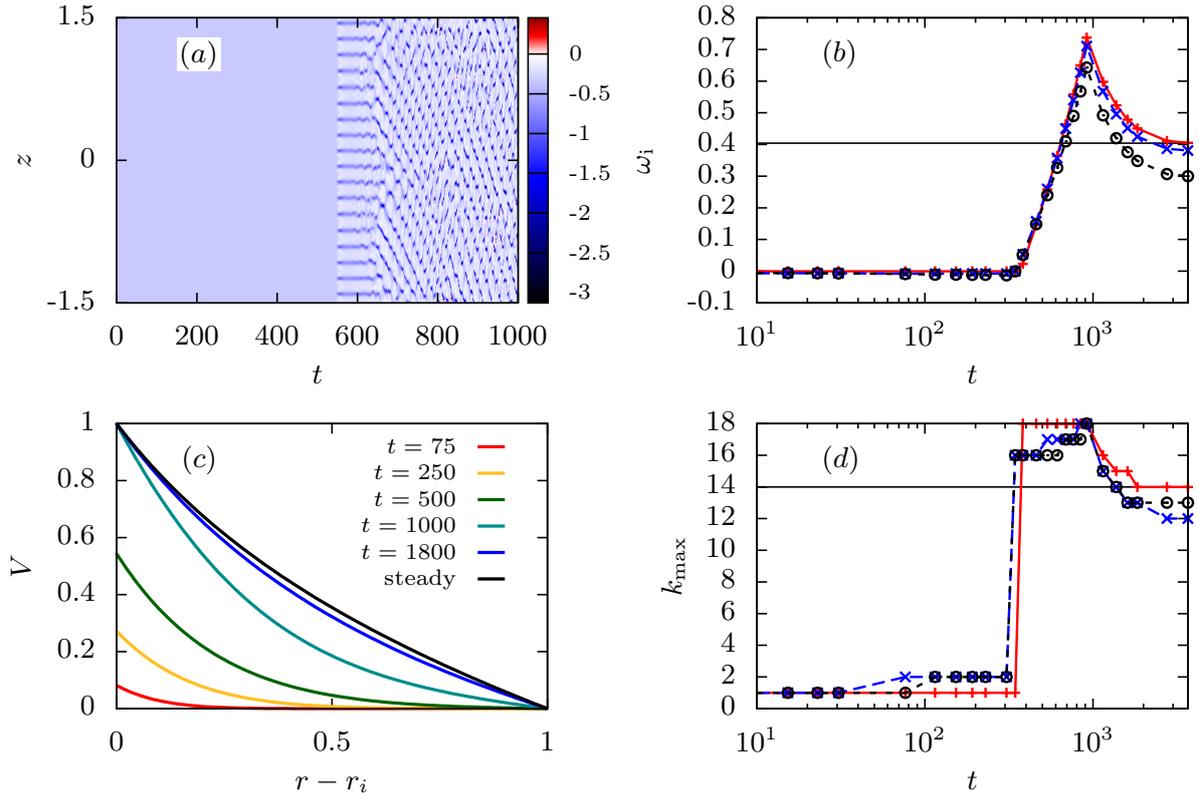

FIGURE 9. Linear stability analysis of quasi-static base flow versus DNS. (*a*) Transiently evolving base flow for a linear spin-up of the inner cylinder over $\tau = 918$ (this choice corresponds to 1h lab time), (*b*) early stages of DNS for run C with $\tau = 918$. Around $t \approx 550$, the dominance of a $(m, k) = (0, 18)$ mode is clearly visible, but helical modes take over rapidly after ($t \gtrsim 650$). In the long term (not shown here) a mixed-ribbon attractor symmetric to that obtained in figure 7(*a*) is obtained. (*c*) Optimal axial wavenumber for $m = 0, 1, 2$ computed on the quasi-static base flow shown in (*a*) and (*d*) associated growth rate (normalized by $\Omega$). The solid line in 7(*c*,*d*) corresponds to the asymptotic value when $t \to \infty$.

For run C, the two different initial conditions lead to the same coherent structure: Span$\{(1, 8), (2, -8)\}$ or its symmetric version Span$\{(1, -8), (2, 8)\}$, despite transient amplification of an $(0, 18)$ mode, as can be seen in figure 9. However, for simulation A, there is not a single pair of robustly attracting states and different initial conditions (same energy in each mode, but with phases synchronized or randomly selected) lead to different attractors: a ribbon Span$\{(2, 6), (2, -6)\}$ in A1 and a mixed-ribbon Span$\{(1, 7), (2, -6)\}$ in A2. These states are visible in figures 4 A1 & A2. Figure 10 shows spatiotemporal diagrams corresponding to these simulations. This time, the vertical derivative of the density field $\tilde{\rho}$ has been averaged over $r$ and $\theta$ so as to show the evolution of the axisymmetric component of the flow. In case A1, there are 12 axisymmetric layers which do not drift vertically, whereas in A2 there are 20 drifting axisymmetric layers. We will come back to this issue in §6.4.

## 6. Discussion

In this section, we will start by reviewing two alternative approaches to try and understand the length scale of the layers. We will first consider the probability distribution function of the density field, which directly connects to the formation of layers in the background stratification profile, to be introduced in §6.1. We will then investigate the relevance of the Ozmidov length scale, to be defined in §6.2, in the selection process. Next, we will revisit literature results in the light of our findings. We will start, in chronological order, with the experiments of Boubnov *et al.* (1995) and Caton *et al.* (1999, 2000) and corresponding computations of Hua *et al.* (1997) in the small-gap, small-$Re < 1800$ regimes. Then finally, we will evaluate the relevance of our numerical results to the more recent experiments of Oglethorpe *et al.* (2013) in the wide-gap and high $Re = O(10^4)$ regime. To simplify notations, we will refer to the aforementioned papers with their associated two letter-two digit acronym: BO95, HU97, CA99, CA00 and OG13.

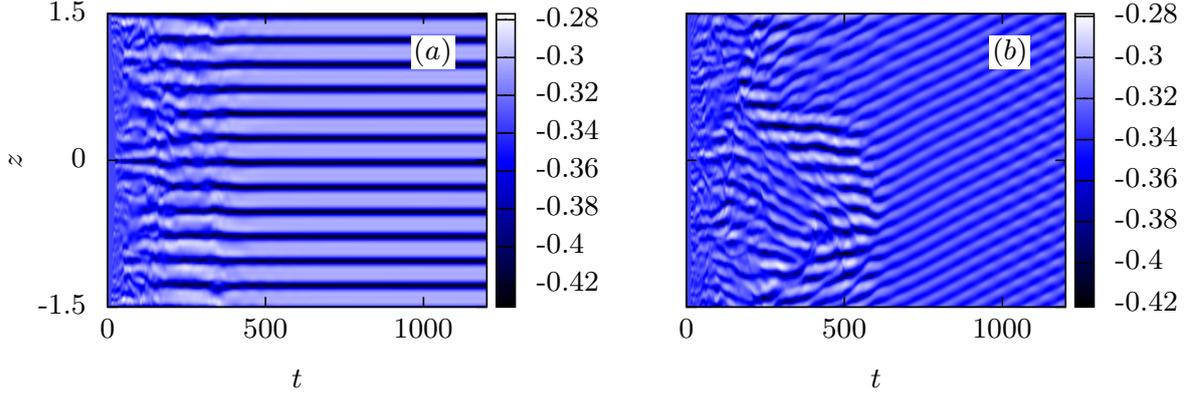

FIGURE 10. Spatiotemporal diagrams of $\langle \partial_z \tilde{\rho} \rangle_S$, where $\langle (.) \rangle_S := \int (.) \, r \, dr \, d\theta$ designates an average over an annular section $S$, for run A and two different initial conditions: (a) all waves of the inital impulse have the same phase, resulting in a ribbon Span$\{(2,6),(2,-6)\}$ producing 12 axisymmetric layers, (b) wave phases are randomized, resulting in a mixed ribbon Span$\{(2,6),(1,-7)\}$ producing 20 axisymmetric layers. Note that the pure ribbon in (a) is a standing wave in $z$ whereas the mixed-ribbon in (b) propagates axially.

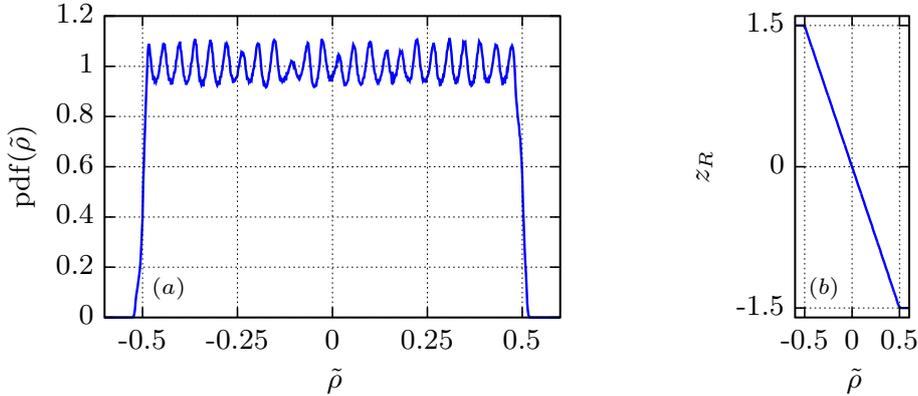

FIGURE 11. (a) Probability distribution function of density field of a snapshot at the end of simulation C. The solid line was obtained by sorting the density field within the computational box $-\Gamma/2 \leqslant z < \Gamma/2$ only. (b) Background density profile $z_R$ computed from pdf using (6.2).

### 6.1. *Layers in the background stratification profile*

To investigate layer formation, we plot snapshots of the probability distribution function of the density field: we expect a higher probability for density values corresponding to well-mixed layers. Figure 11(a) gives a typical example obtained from a mixed-ribbon (run C). The first striking feature of the pdf is, as expected, the presence of peaks. There are 24 peaks, which matches exactly with $k_{\text{axi}}$ as defined in (5.4). It is therefore tempting to associate the peaks to the signature of the axisymmetric subspace of the flow. However, correlation does not imply causation, and we found that the number of peaks is in fact the signature of the two dominant helical modes of the flow instead. Indeed, computing the pdf of synthetic two-dimensional density fields of the general form

$$\tilde{\rho}(\theta, z) = -\frac{z}{\Gamma} + K[\cos(m_1\theta + k_1 k_0 z) + \cos(m_2\theta + k_2 k_0 z)], \quad (6.1)$$

with $K$ some positive amplitude factor (such that the density profile is nowhere 'unstably' stratified, in the sense $\partial_z \tilde{\rho} > 0$) and $m_1 m_2 \neq 0$, lead to similar types of pdf, with a number of peaks corresponding exactly to $k_{\mathrm{axi}}$, despite the absence of axisymmetric modes. The dominant helical modes interact through the highly nonlinear pdf operator, and the emergence of $k_{\mathrm{axi}}$ in figure 11 is not necessarily a consequence of the inertial term in the Navier–Stokes equations.

We note in passing that whereas a single axisymmetric mode can induce a change in the pdf, a single non-axisymmetric mode cannot. At least two linearly independent non-axisymmetric modes are required to modify the pdf, and therefore the background stratification. Indeed, the two modes can have the same handedness (same sign of $mk$), but if they are also linearly dependent (same pitch $m/(kk_0)$), which is the case in saturated helical branches (dominant mode plus harmonics of lower amplitude), the pdf is not modified.

The second remarkable feature of the pdf is the low amplitude of the variations around 1. This property was verified for all simulations A–G, including the most 'disordered' ones (F and G in particular, see figure 6). This is likely to be a Schmidt number effect, as will be seen in §6.3. The direct consequence of this observation is the absence of significant layering in the background stratification profile $z_R$, as evident from figure 11($b$). We recall that $z_R$ corresponds to the height associated with a fluid particle after adiabatic sorting of the density field into its stage of minimum gravitational potential energy, and can be obtained at any time by integration of the pdf (Tseng & Ferziger 2001):

$$\mathrm{d} z_R|_t = -\Gamma \, \mathrm{pdf}(\tilde{\rho}, t) \mathrm{d}\tilde{\rho}, \quad z_R(\max_{\mathcal{V}} \tilde{\rho}) = -\frac{\Gamma}{2}. \quad (6.2)$$

In the rest of this paper, we will distinguish the number of (localized) 'dynamic' layers, corresponding to $k_1 \approx k_2$ and visible in figures 4 and 5, from the the number of layers appearing in the background stratification $z_R$, which we call 'static'.

### 6.2. *Layers and turbulence: Ozmidov length scale*

In the context of stratified turbulence, the Ozmidov length scale

$$l_O := \sqrt{\epsilon/N^3}, \quad (6.3)$$

based on the kinetic energy dissipation $\epsilon$, gives an estimate of the smallest scale influenced by buoyancy. This quantity may therefore be considered a candidate for predicting the vertical scale of the layers. The kinetic energy dissipation is defined as the volume integral of $\tau_{ij}\partial_j u_i$, where $\tau_{ij}$ is the viscous stress tensor, expressed as $\tau_{ij} = 1/Re(\partial_i u_j + \partial_j u_i)$ if $\epsilon$ is normalized by $\rho_0(r_i\Omega)^3\Delta r^2$. Table 1 gives the value of $l_O$ for all our simulations: it is of the order of 1/100 gap for A–G, which is one order of magnitude smaller than the 'dynamic layer' depth $\Gamma/k_1 \approx \Gamma/k_2$ or the 'static layer' depth $\Gamma/k_{\mathrm{axi}}$: the flow does not dissipate enough energy for this length scale to be relevant, even at $Re = 10^4$ (simulation E).

To better understand this mismatch, we introduce the buoyancy Reynolds number

$$Re_b := \frac{\epsilon}{\nu N^2}, \quad (6.4)$$

which is linked to the Ozmidov scale by the relation $Re_b = (l_O/l_K)^{4/3}$, where $l_K := (\nu^3/\epsilon)^{1/4}$ is the Kolmogorov scale. In all simulations A–G, the buoyancy Reynolds number is of order 1, except for the smaller gap case $\eta = 0.625$ where it reaches the value of 12 (see table 1). This means that there is no significant separation of length scales between $l_O$ and $l_K$, and as a result, the regime of strongly stratified turbulence is not achieved (it is usually considered that $Re_b$ should be larger than at least 20 (Smyth & Moum 2000)). It is therefore not suprising to find $l_O$ to be irrelevant to predicting the vertical scale of the layers.

Simulation CT however, to be discussed in the next section, reached a high buoyancy Reynolds number of $Re_b = 123$, and therefore could be considered to belong to the strongly stratified regime. But even then, the buoyancy Reynolds number understimates layer depth, which confirms the idea that the layer formation mechanism may not be an inherently turbulent process in STC, at least up to $Sc = 16$.

### 6.3. *Layer formation in BO95's experiments*

In BO95, the authors established a phase diagram with 6 different flow regimes, plus the laminar Couette solution, in their $\eta = 0.769$ apparatus. CA00 amended this diagram by suppressing the transition zone 'ST', which could no longer be found in their enhanced apparatus. HU97 then ran direct numerical simulations at a fixed value of the Grashof number $G = N^2 d^4/\nu^2 = 22430$, corresponding to $N = 1.04\,(\text{rad/s})$ in the apparatus of BO95, for which $\Delta r = 12\,\text{mm}$. HU97 considered three values of $Re = 245, 480, 816$, respectively falling in the S ('stratified vortices'),T ('Taylor vortices') and CT ('Taylor vortices coupled in pairs') regions of BO95's diagram. They used a Schmidt number value of $Sc = 16$, based on their finding that the (axisymmetric) linear instability threshold does not vary significantly beyond that value.

In this section, we revisit these computations and therefore use the same values of $\eta$, $G$, $Re$ and $Sc$ as HU97 (the corresponding values of $Ri$ are obtained from the relation $Ri = [G/Re^2][\eta^2/(1-\eta)^2]$). But while in the original experiment of BO95 the aspect ratio $\Gamma = 52$ was quite large, we decided to choose a domain length 16 times smaller for the DNS: $\Gamma = 3.25$. This is still larger than the domains considered by HU97, where $\Gamma < 2.8$.

#### 6.3.1. *Phase diagram of BO95: S, T and CT regimes*

Figure 12 shows our results, for impulsively started simulations in the S and CT regimes, in the form of snapshots of the density field in a meridional plane and of the background stratification profile. Figure 12(*a*) is highly reminiscent of the dye visualization of CA99 (figure 1(*b*) therein), giving us confidence in the fact we have computed a flow very close to the experiment. Despite the low $Re_b \approx 5.4$ and the fact that the flow is laminar, we do observe layering of the background stratification profile, confirming that the separation of scales induced by $Sc \gg 1$ is the key condition for 'static' layer formation. The spectrum of perturbation kinetic energy shows that the coherent structure is a pure ribbon Span$\{(3,3),(3,-3)\}$, with our choice of $\Gamma$. The number of 'static layers' in figure 12(*b*) is 6, which corresponds to $k_{\text{axi}}$ from equation (5.4) and is therefore consistent with our analysis in §6.1: the number of 'static' layers is determined by nonlinear interaction of the dominant modes through the pdf operator.

The density field in the CT simulation is highly disordered, leading to a background stratification which is far from linear. One density layer is half the size of the other, which is consistent with the description of BO95. The spectrum of kinetic energy is complex, but clearly involves energetic contributions from modes $(0, \pm 1)$ and $(0, \pm 2)$, as well as powerful helical modes. The layers in CT seem much better mixed than in any other simulation, and this can be quantified through the density standard deviation $\sigma^{1/2}$, which turns out to be one order of magnitude larger in CT, compared to A–G and S.

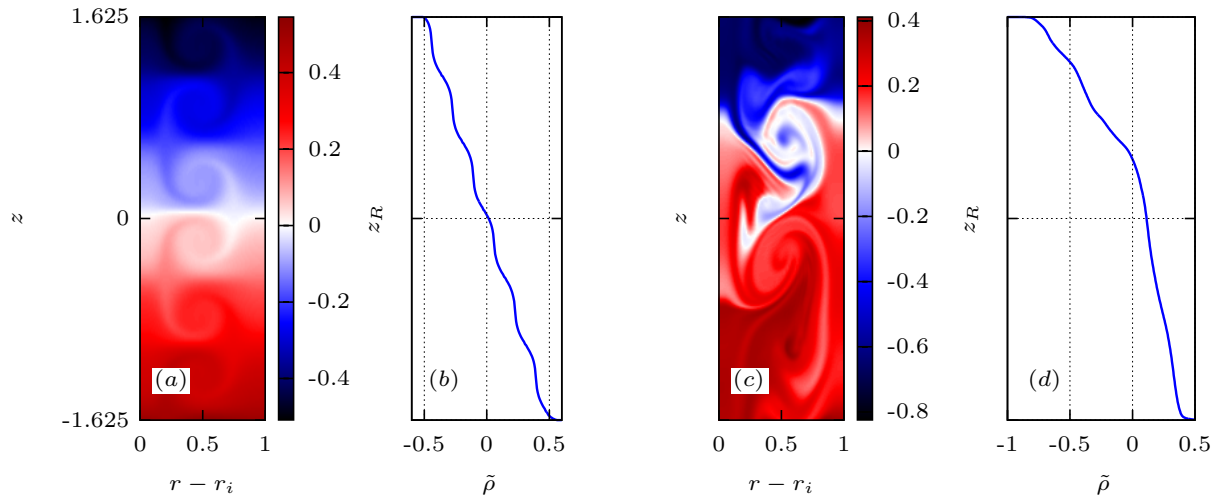

FIGURE 12. Snapshots of the density field $\tilde{\rho}$ for the (a) S and (c) CT regimes of BO95. (b, d) Corresponding background stratification profiles.

We note the clear correlation between large $\sigma^{1/2}$ and large $Re_b$: the more vigorous the turbulence, the more the density field deviates from the initially linear profile.

Finally, we did not find an axisymmetric steady solution, using the parameters of the T regime. We suspect that this is related to the impulsive start of the simulation, leading to a different flow state (HU97 computed axisymmetric steady solutions in small domains, which they then perturbed with azimuthal noise).

6.3.2. *Nature of the bifurcation to the non-axisymmetric S regime*

While BO95 observed a primary bifurcation to the non-axisymmetric state S, HU97 found a transition to an axisymmetric oscillatory state instead. The latter authors also found the S branch in their simulations, but only at a Reynolds number slightly larger than the critical $Re$ of the primary bifurcation. CA99 and CA00 later proposed that the non-axisymmetric 'vortex branch' S was created by a subcritical bifurcation, becoming nonlinearly attracting at a saddle-node bifurcation occurring slightly above the critical $Re$ of the primary bifurcation associated with the axisymmetric branch. The hysteresis observed experimentally in switching from one branch to the other as $Re$ was varied up or down was interpreted as a global bifurcation between two limit cycles.

While CA99 and CA00 used linear stability analysis, they did not consider non-axisymmetric modes. In figure 13(a), we present the complete analysis, including modes $m = 0, 1, 2$ for $Sc = 730$ and the whole range of buoyancy frequencies/Grashof numbers considered by these authors. We show that the critical Reynolds numbers between the different mode numbers are extremely close, especially between $m = 0$ and 1. Inputting the critical modes for $m = 1$ at $G = 22430$ into the DNS code (on top of the base flow) and restricting $\Gamma$ to fit only one axial wavelength, we showed that the corresponding helical and ribbon branches both bifurcate supercritically (see figure 13(b)). Note that this nonlinear computation has been done with the experimental value of the Schmidt number, i.e. $Sc = 730$. This suggests that the observation of $m = 1$ or $m = 0$ at onset, depending on initial conditions, may simply be the consequence of the proximity of their critical $Re$. It is also tempting to reinterpret a conclusion of HU97 in the light of this finding: 'the axial scale selection [...] is primarily determined by the axisymmetric part of the flow, while the azimuthally dependent component plays a lesser role'. In fact, the critical axial wavenumbers are nearly indistinguisable between $m = 0$ and $m = 1$, which would explain in turn why the axisymmetric and non-axisymmetric nonlinear branches have similar wavelengths.

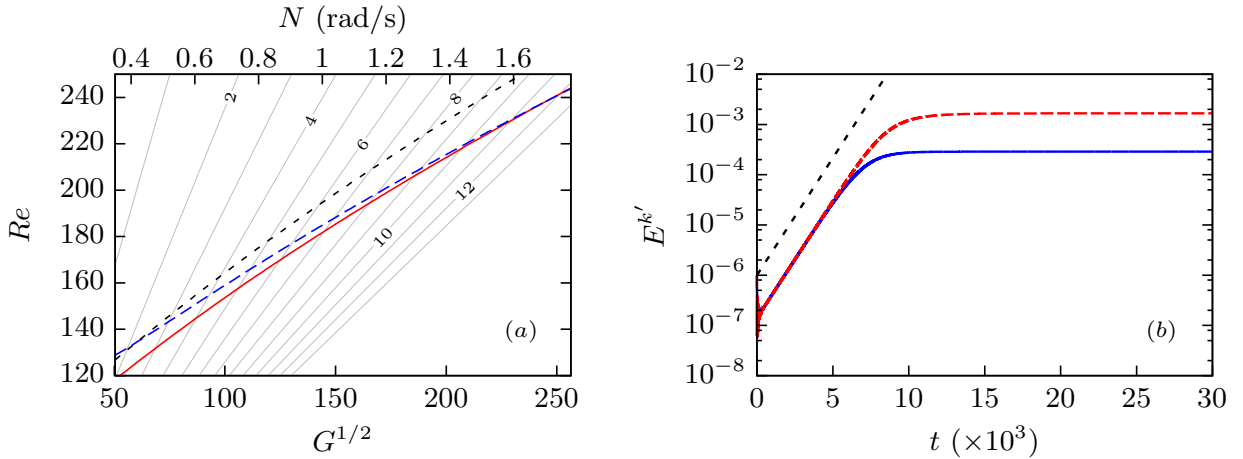

FIGURE 13. (*a*) Critical $Re$ versus square root of the Grashof number $G$ for $\eta = 0.769$ and $Sc = 730$, as in BO95, CA99, CA00. Contours of constant $Ri$ are indicated with solid grey lines. The range of $G$ corresponds to the range of buoyancy frequencies considered by the authors, and is indicated on the top horizontal axis. Solid line $m = 0$ (oscillatory), dashed line $m = 1$, dotted line $m = 2$. (*b*) Perturbation kinetic energy $E^{k'}$ versus time, showing the supercritical bifurcations of the helical (solid line) and ribbon (dashed line) branches created by mode $m = 1$ at $G = 22430, Re = 189$, right above its instability threshold $Re_c = 188.1$. The dotted line corresponds to the prediction from linear analysis $E^{k'} \propto \exp(2\omega_i t)$.

Remarkably, the critical $m = 1$ mode found in figure 13 propagates against the flow, as $\omega_r/m < 0$, which is contrary to what was found for $\eta = 0.417$ in §4. This means that such mode cannot be taking energy from the base flow through a critical layer mechanism, and is yet another illustration of a non-axisymmetric instability which is not directly related to the stratorotational mechanism of Park & Billant (2013).

One could argue however, that the S state does not correspond to either the ribbon or the helical branches created by the critical mode from linear analysis. Indeed, the flow found in the DNS of the S regime at $Sc = 16$ (figure 12(*a*)) corresponds in fact to a ribbon created by $m = 3$ modes, and propagating in the same direction as the base flow. At $Sc = 16$, the helical modes which create this structure bifurcate at $Re = 204.37$, and we have checked that this bifurcation is also supercritical. Therefore, it seems unlikely that the bifurcation of the vortex branch be supercritical, even though we do not exactly know which attractor S really was in the initial experiment.

### 6.4. *Layer formation in OG13's experiments*

OG13 studied STC flow at much larger Reynolds numbers $Re = O(10^4)$ than previous authors, in order to focus on turbulent mixing. Their apparatus had a very large gap varying in the range $\eta \in \{0.195, 0.389, 0.584\}$ and a small aspect ratio $1.2 < \Gamma < 4.6$. The authors studied two types of initial stratification profiles: linear or with five homogeneous layers. We will only focus on the initially linear experiments here. For these, the authors were able to successfully fit their data for the depth of homogeneous density layers using the scalings derived by BO95 with an energetic argument (assuming an axisymmetric steady flow). In this section, our purpose is to evaluate whether these results are captured by our low-$Sc$ DNS and to which extent the present study can illuminate the layer formation mechanism in experimental conditions.

The most noticeable difference between our results and that of OG13 lies in the structure of the layers: well-mixed, with sharp density interfaces in the experiments versus highly three-dimensional structures with radially and azimuthally localized interfaces in the DNS. Given the trend in our simulations at respectively $Sc = 1, 10$ and $16$, we make the assumption that the smoothness of the interfaces far from the inner cylinder is a low Schmidt number effect. However, the mismatch in $Sc$ does not constitute an obvious explanation for the apparent difference in the azimuthal structure of the layers. In the spatiotemporal diagrams of OG13, interfaces are highlighted by thick solid lines, which are approximately horizontal in the $(t, z)$-plane (the slow downward motion of the interfaces is accounted for by fluid suction by the probe). These continuous horizontal lines suggest that the layers are well-mixed and axisymmetric. But direct numerical simulations yield non-axisymmetric coherent structures, leading to the apparent 'flashing' of interfaces as the structure rotates around the annulus. These coherent structures are created by saturation of linear instability modes which seem weakly sensitive to the Schmidt number, hence could also appear at large $Sc = 700$.

To further understand this paradox, a new series of experiments were performed, with a similar protocol as OG13, but with a radius ratio of $\eta = 0.417$ and an aspect ratio of $\Gamma = 3$ (initially motivating the choices of $\eta$ and $\Gamma$ in the DNS). In these impulsively started experiments, shadowgraphs were used in order to visualise flow structures. The outer cylinder was made of perspex and was transparent whilst the inner cylinder was painted matt white. A slide projector was positioned 2-3 m away from the apparatus to illuminate the flow. The refractive index variations with the fluid, produced by density differences (caused by variations in NaCl concentration), resulted in shadowgraph images being projected onto the white inner cylinder. These shadowgraph images were subsequently recorded by a digital camera. Additionally, conductivity measurements were taken with a fixed probe located at $r - r_i = 36\%$ (and fixed $\theta$), translating vertically at an enhanced rate compared to OG13. The probe traversed a large proportion (¿80%) of the full depth of the fluid column in approximately 3-4 s with a period between probe measurements of 20-40 s. The time taken to traverse in our new, high-speed measurements is on the order of a rotation period of the inner cylinder which is must faster than those previously possibly in the study of OG13, where the probe took approximately 2 min to traverse the fluid, i.e. $O(10)$ rotation periods. The spatiotemporal shadowgraph in figure 14 reveals the presence of powerful nonlinear waves, with a periodic disappearance of interfaces highly reminiscent of our DNS results (in particular run D at comparable $Re$ and $Ri$). These structures are now also visible in the conductivity measurements, as can be seen in figure 15. This suggests that the limited vertical speed of the probe in OG13 might have acted as a filter of unsteady/non-axisymmetric motions. Ribbon structures were already observed in STC by Le Bars & Le Gal (2007), but at much lower $Re < 1200$ and with a much smaller gap $\eta = 0.8$. The pattern shown here appears to be a turbulent version of the ribbon, similar to turbulent Taylor vortices in the unstratified case Brauckmann & Eckhardt (2013). We note in passing that Withjack & Chen (1974) were probably the first to observe ribbon-type structures in an experiment (even before the name of the structure was coined by Demay & Iooss (1984) in their theoretical analysis of counter-rotating unstratified Taylor–Couette flow): 'with counterrotating cylinders, the instabilities appear as regularly spaced vortices which, for the most part, are neither symmetric Taylor vortices nor simple spirals. In addition, these vortices rotate as a whole at a speed generally smaller than that of the inner cylinder'.

Our numerical results seem qualitatively compatible with figures 14 and 15, but can we, in general, make quantitative prediction of the temporal and length scales? By applying Fourier transform to spatiotemporal diagrams, we retrieved the axial wavenumbers of the interacting helices and their temporal frequencies. Results are reported in 2, in the form of the number of axisymmetric, or 'static', layers $k_{\text{axi}}$ defined in (5.4) and the mean azimuthal phase speed $\omega^\star$ defined in (5.5). Since the experiments all seemed to be dominated by $m = 1$ ribbon branches, $\omega^\star$ simply corresponds to the dominant frequency of the flow. We chose to present the number of axisymmetric layers $k_{\text{axi}}$ rather than the number of 'dynamic layers' $k_1 = k_2$ (wavenumbers of the dominant helical modes) in order to match OG13's characterisation of the layers. There is reasonable agreement overall: $0.17 < \omega^\star < 0.19$ and $17 \leqslant k_{\text{axi}} \leqslant 27$ in experiments versus $0.19 < \omega^\star < 0.3$ and $16 \leqslant k_{\text{axi}} \leqslant 24$ in the DNS for the same range of $Ri$.

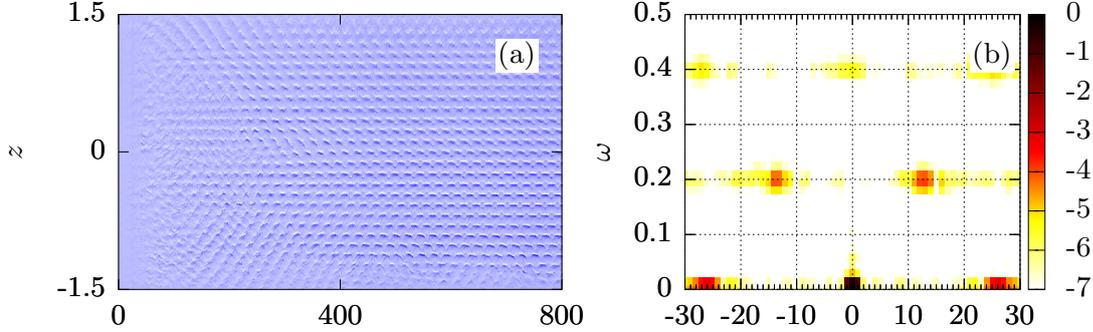

FIGURE 14. (*a*) Spatiotemporal shadowgraph (in false colors) for $Ri = 10.48$, $Re = 7000$, following an impulsive start. (*b*) Corresponding two-dimensional Fourier transform with $\omega$ normalised with the rotation rate $\Omega$.

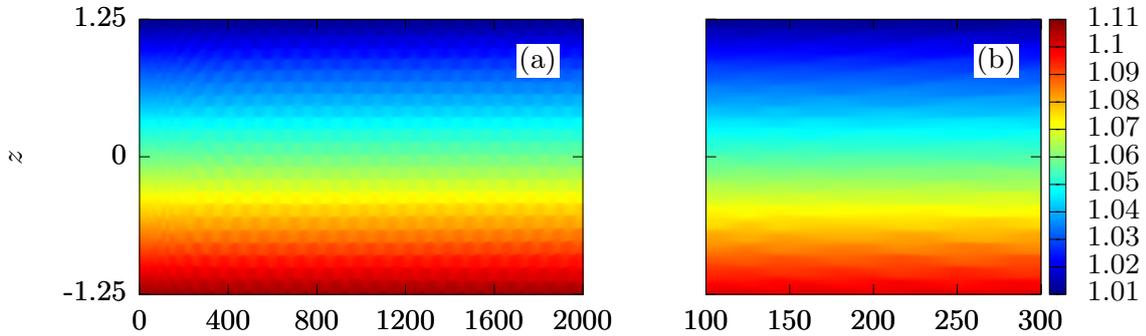

FIGURE 15. *For now, this figure is just a placeholder. Plot drhodz instead to be able to count interfaces in (b)? Get rid of the long-time diagram and keep only the zoom?* Spatiotemporal conductivity diagram for the same experiment as in figure 14. The probe is located at $r - r_i = 0.36$.

However, whereas the number of layers increases almost monotonically with $Ri$ in the experiment, it sometimes evolves non-monotonically in the DNS. This may be explained by the switch of coherent structure from a mixed-ribbon at $Ri = 2$ and 3 to a pure ribbon at $Ri = 10$. The values of $k_1$ and $k_2$ increase steadily with $Ri$, so the number of dynamic layers does increase, but the change in the nature of the coherent structure causes a sudden drop in the number of 'static layers' $k_\mathrm{axi}$. Such a dramatic effect of the nature of the coherent structure has already been observed in simulation A (section §5.3): a minor change in initial conditions lead to either $k_\mathrm{axi} = 12$ or $k_\mathrm{axi} = 20$, depending whether mode $(2, 6)$ was interacting with $(2, -6)$ or $(1, -7)$. But despite this leap in $k_\mathrm{axi}$, the axial wavenumbers of interacting helices $k_1$ and $k_2$ are approximately the same between A1 and A2.

| $Ri$ | $Re$ | $k_\mathrm{axi}$ | $\omega^\star$ |
|---|---|---|---|
| 2.39 | 14000 | 17 | 0.17 |
| 2.60 | 14000 | 18 | 0.16 |
| 2.69 | 7000 | 21 | 0.18 |
| 2.79 | 10500 | 18 | 0.17 |
| 4.68 | 10500 | 22 | 0.18 |
| 6.00 | 7000 | 24 | 0.19 |
| 10.48 | 7000 | 27 | 0.19 |

TABLE 2. New series of experiments with $\eta = 0.417$, $\Gamma = 3 \pm 2\%$ and $2 < Ri < 10$, as in the DNS, using spatiotemporal shadowgraph to determine the number of axisymmetric layers $k_\mathrm{axi}$ and the dominant frequency $\omega^\star$ Why does $k_{axi}$ goes from 21 to 18 as $Ri$ increases from 2.69 to 2.79?).

Additionally, end-plates add a reflection symmetry to the system which is not captured by the DNS. This may potentially constrain the choice of coherent structures further. Another hint in that direction is the sudden changes in the number of layers reported by OG13, as top and bottom layers grow in time. OG13 interpreted these events as overturning of interfaces, but this may also correspond to Eckhaus bifurcations between different coherent structures caused by evolving end-effects. In any case, the high sensitivity of $k_\mathrm{axi}$ to the nature of the coherent structure makes it hard to make quantitative comparisons between DNS and experiments.

Finally, we are left with an open question regarding the interpretation of the length scale selection mechanism in these experiments. In the DNS, the layer depth seems to be fixed by the coherent vortical motions, which are nonlinearly selected from the set of linearly unstable wavenumbers. Unfortunately, there currently exists no theoretical criterion to anticipate the dominant wavenumbers of a coherent structure in the fully nonlinear regime (weakly nonlinear analysis could potentially be applied at slightly supercritical Reynolds numbers $Re = O(10^2)$), and the use of an energetic argument of the form of BO95 and OG13 may potentially be inappropriate in that context. Indeed, the steady version of Bernoulli's theorem does not apply to rotating structures like ribbons. And even in the case of an axisymmetric structure like a Taylor vortex, the potential energy of fluid particles would not change along trajectories in the unstratified case, making energetic arguments inapplicable in the passive-scalar limit $Ri \to 0$. Future computations at higher $Sc$ number will surely help settle the debate between the two proposed length scale mechanisms.

## 7. Conclusions

In this paper, we have studied numerically the linear and nonlinear dynamics of the Taylor–Couette flow stratified in the axial direction (STC), with the aim of understanding the mechanism leading to layer formation in experiments.

Using linear stability analysis, we showed that the critical perturbation is always non-axisymmetric for our reference radius ratio $\eta = 0.417$, when the Richardson number $Ri$ is of order 1. Despite the fact that the flow is centrifugally unstable according to Rayleigh's criterion, linear axisymmetric instabilities only dominate at very large Reynolds number, such that the bifurcation is controlled by helical rather than toroidal perturbations. The main effect of the Schmidt number is to allow the dominance of unsteady $m = 0$ perturbations for very small $Ri \ll 1$, only if $Sc$ is large enough. Otherwise, the effect of the Schmidt number on the linear dynamics remains small for $1 \leqslant Sc \leqslant 700$. Therefore, if linear theory has some relevance to the nonlinear dynamics, then it is appropriate to carry numerical simulations at low $Sc$, for $Ri$ of order 1.

Direct numerical simulations at $Sc = 1$ and $Sc = 10$ for $Re$ up to $10^4$ showed that the nonlinear dynamics is always dominated by coherent structures formed by the interaction of a pair of unstable helical modes of opposite handedness. The structures formed are called (mixed)-ribbon and (mixed)-cross-spirals, depending whether the two helices are mirror-symmetric and whether they have comparable amplitude. Although the two leading modes are both linearly unstable, the coherent structures formed by their interaction are strongly nonlinear, therefore the dominant wavenumbers cannot be predicted by linear theory. Moreover, the selection of the coherent structure may be sensitive to initial conditions. At large enough $Re$, the coherent structures correspond to saddles in phase space, organising the weakly turbulent dynamics.

The large coherent vortical motions overturn the density field, leading to the formation of layers and sharp density interfaces at the location of the radial jets bounding the vortices. However, given the strongly inhomogeneous nature of the coherent structures, the resulting 'dynamic' layers are not axisymmetric and the interfaces are sharper near the walls than in the interior. The strong activity near the walls is reminiscent of the structure of the helical linear modes, which themselves recall the structure of stratorotational instability modes (caused by the interaction between inertia-gravity waves localized at each wall). In the case of a ribbon, the two interacting nonlinear waves propagate at the same speed in opposite axial direction, leading to a standing pattern in $z$. However, both waves rotate azimuthally in the same direction, at approximately the mean angular velocity of the flow, leading to the apparent 'flashing' of long-lived density interfaces, periodically disappearing and reappearing in spatiotemporal diagrams.

Surprisingly, the number of layers observed instantaneously, at a given azimuthal and radial position, does not match with the number of peaks in the probability distribution function of the density field. Instead, this number seems to correspond to the wavenumber $k_{\text{axi}}$ generating the axisymmetric subspace of the coherent structure. However, the same number of peaks appears in the pdf even if axisymmetric components of the density field are completely filtered out. This means that the peaks can be created by interactions of the dominant helical modes directly through the nonlinear pdf operator and not necessarily through the nonlinear term in the Navier–Stokes equations. In any case, this number corresponds to the layers appearing in the adiabatically sorted density profile, or background stratification. We therefore distinguish the number of 'static' layers, given by $k_{\text{axi}}$, from the number of 'dynamic', localized layers, given by the wavenumbers of the two dominant modes.

In the $Sc = 1$ and $Sc = 10$ simulations, 'static layering' is weak as the deviation of the background stratification from its initially linear profile is negligible, even at $Re = 10^4$. This is likely a $Sc$ number effect, as a visible impact already occurs at $Re = O(10^2)$ for a $Sc = 16$ simulation. Moreover, layers appear to survive after stopping experiments with salty water, which means that strong static layering must occur at $Sc = 700$. This is a fundamental qualitative difference between our numerics and experiments. However, spatiotemporal shadowgraph confirm the existence of powerful nonlinear waves in STC, at $Re = O(10^3 - 10^4)$. The frequency of the dominant waves and the number of axisymmetric layers of these coherent structures is in reasonable agreement with the DNS, which suggests that the length scale selection mechanism was correctly captured qualitatively by our analysis, despite the $Sc$-number limitation. However, more precise predictions of the layer depth may be hard to achieve with an axially periodic model of STC, even if the $Sc$ mismatch could be lifted. Indeed, sensitivity to initial conditions and an additional symmetry imposed by the end-plates in practice may lead to the selection of a different coherent structure in practice.

In the scenario described in this paper, the role of turbulence seems secondary in the length scale selection process, even at $Re = 10^4$ for low $Sc$. The Ozmidov length scale largely underestimates the size of the layers, but the buoyancy Reynolds number $Re_b$ is not large enough is most of our simulations to consider them in the regime of 'strongly stratified turbulence' Brethouwer *et al.* (2007). However, one of our simulations with $Sc = 16$ did reach $Re_b = O(10^2)$ and it apparently lead to enhanced static layering and a much larger density variance. But the length scale selection still appears to be based on an underlying coherent structure.

We have also revisited numerical and experimental results from the literature. First we have reproduced the computations of Hua *et al.* (1997) at $Pr = 16$ in the S, T and CT regions of the experimental diagram of Boubnov *et al.* (1995), but in a larger domain and starting simulations impulsively. Contrary to Boubnov *et al.* (1995) and Hua *et al.* (1997), we found a non-axisymmetric and time-dependent flow in the T zone, which is likely to be the signature of sensitivity to initial conditions. The S regime appears to be a ribbon branch caused by $m = 3$ helical modes. This 'vortex branch' described by Caton *et al.* (1999, 2000) was initially found to bifurcate first from the base flow in the early experiments of Boubnov *et al.* (1995). But subsequent simulations (Hua *et al.* 1997) and careful experiments (Caton *et al.* 1999, 2000) found a primary bifurcation to an axisymmetric 'wave branch' instead, leading to a contradiction with Boubnov *et al.* (1995). Caton *et al.* (1999, 2000) proposed a novel bifurcation scenario to explain this discrepancy, where the wave and vortex branches would bifurcate respectively supercritically and subcritically from the base flow, but where the latter branch would become stable through a saddle-node bifurcation almost at the point where the wave branch bifurcates. We propose instead that the two branches both bifurcate supercritically, and almost simultaneously. Indeed, using linear stability analysis, we showed that the instability thresholds of the different modes are very close, although the axisymmetric mode grows slightly faster. Using DNS, we established the supercriticality of the bifurcation of the ribbon branch found in the DNS at $Sc = 16$. We also checked that the ribbon branch created by the critical $m = 1$ mode at $Sc = 730$ also bifurcates supercritically. It is therefore plausible that the discrepancy between the initial observations made by Boubnov *et al.* (1995) and subsequent authors is simply due to the proximity of the two branches in the bifurcation diagram. More importantly for us, this shows once again that non-axisymmetric linearly unstable modes can play a major role in the non-linear dynamics, even when the critical mode is axisymmetric.

Our large-$Re$, large-gap experiments in a regime close to that of Oglethorpe *et al.* (2013) seem to indicate that the layer depth measured by these authors likely corresponded to the depth of axisymmetrically averaged rather than instantaneous structures, because of artificial filtering by a slowly moving probe. Non-axisymmetric coherent structures would have therefore come unnoticed without the use of shadowgraphy. Unfortunately, the use of an energetic argument to predict layer depth becomes problematic if the flow was really dominated by non-axisymmetric, unsteady structures like ribbons.

Despite encouraging progress in understanding former experiments, we reiterate a major limitation of our simulations, which seem unable to fully capture layer formation because of the strong impact of the Schmidt number. Future numerical work will therefore be targeted at large $Sc$ simulations, using high resolution DNS to check whether the mechanism uncovered in this study still apply. We expect a strong impact on the background stratification, but only a weak effect on layer depth and dominant frequency selection. From an experimental viewpoint, it would be interesting to obtain density measurements in meridional and horizontal slices, in order to resolve the three-dimensional structure of the layers, which remains unknown in experiments. The implications of the identification of powerful nonlinear waves on the buoyancy flux measured by Oglethorpe *et al.* (2013) is discussed separately in Part 2 (Leclercq *et al.* ????*b*).

## Acknowledgements

We are indebted to Liang Shi, Markus Rampp, José-Manuel Lopez, Björn Hof and Marc Avila for sharing with us their excellent DNS code (Shi *et al.* 2015). C. L. gratefully acknowledges fruitful collaboration with R. R. Kerswell. Sebastian Altmeyer and John R. Taylor are also thanked for enlightening discussions. This work has been supported by the EPSRC, under grant EP/K034529/1.